\documentstyle[aps,pre,epsf]{revtex}

\begin{document} 
\draft
\author{C. L. Emmott}

\address{Department of Physics and Astronomy, University of Manchester, 
Manchester, M13 9PL, UK.}

\title{Perturbative Corrections to the Ohta-Jasnow-Kawasaki Theory of
Phase-Ordering Dynamics}

\date{\today}
\maketitle

\begin{abstract}
A perturbation expansion is considered about the Ohta-Jasnow-Kawasaki
theory of phase-ordering dynamics; the non-linear terms neglected in
the OJK calculation are reinstated and treated as a perturbation to
the linearised equation.  The first order correction term to the
pair correlation function is calculated in the large-$d$ limit and
found to be of order $1/d^2$.
\end{abstract}


%\twocolumn
%\vskip.2pc]
%\narrowtext

%\newpage

%+++++++++++++++++++++++++++++++++++++++++++++++++++++++++++
%+++++++++++++++++++++++++++++++++++++++++++++++++++++++++++

\section{Introduction}
\label{intro}       

When a system is quenched from a high-temperature, homogeneous phase,
into a two-phase region, domains of the equilibrium phases form and
evolve with time.  If the order parameter is non-conserved, then the
coarsening dynamics are modelled by the time-dependent
Ginzburg-Landau (TDGL) equation \cite{Review}.

In one dimension this system is exactly soluble
\cite{Nagai,Derrida,Bray90b}, but due to the non-linear nature of
the TDGL equation, no exact solutions are available for a general
number of dimensions so we must rely on approximate
theories.  There are several approximate theories which describe the
coarsening dynamics of this system.  However,  
 they all rely on a similar approach 
in which the order parameter, $\phi$, is replaced by a smoothly
 varying auxiliary field, $m$, which
has the same sign as $\phi$, and whose zeros define the domain walls.
The equation of motion (TDGL) is then recast in terms of this
auxiliary field and various approximations are made in order to make
 the resulting
 equation soluble.
In the Ohta-Jasnow-Kawasaki (OJK) \cite{OJK} theory,
 this is achieved by replacing the non-linear terms
 by their spherical average, thus linearising the equation for the
auxiliary field .

Comparison of the OJK results with simulation data \cite{Humayun92c}
 have shown that
this theory gives a very good description of the system.  However, the
approximation is uncontrolled and it is unclear why this approach
gives such accurate predictions. Furthermore,
 Blundell {\it et. al.} \cite{Blundell} have pointed out that by
 plotting $(1-C_{\phi})$
 against $1/C_{\phi^2}$,
where 
\begin{eqnarray}
C_{\phi}&=& \langle\phi({\bf x}+{\bf r},t)\phi({\bf x},t)\rangle, \\
C_{\phi^2}&=& {\langle (1-\phi^2({\bf x}+{\bf r},t)) (1-\phi^2({\bf
x},t))\rangle \over
\langle 1-\phi^2({\bf x}+{\bf r},t)\rangle
\langle 1-\phi^2({\bf x},t)\rangle},
\end{eqnarray}
 thereby removing any adjustable parameters,
 the OJK theory does not follow the simulation data as closely as
 previous comparisons, using only $C_\phi$, suggested.  

An obvious way to examine this
 approximation is to treat the
neglected non-linear terms as a perturbation to the linearized
 equation of motion, and this is exactly the approach taken in this
 paper.

  This
calculation has two main advantages:
firstly, in OJK theory the initial conditions are conventionally taken
to be Gaussian; since the evolution equation is linear, this
will ensure that the distribution for the auxiliary field at later times
is also Gaussian.  This feature of OJK theory 
 has been questioned by Yeung {\it et. al.} \cite{Yeung94}. 
Their simulations, which explicitly calculate
the auxiliary field distribution, give results which are not
 exactly Gaussian, particularly at small values of the auxiliary 
field.
We note that recently, in a series of papers, Mazenko 
\cite{Mazenko1,Mazenko97}, 
and Mazenko and Wickham \cite{Mazenko2}, have
 presented an approximate theory
which goes beyond the Gaussian distribution.

Secondly, it has been proposed by both Bray and Humayun \cite{Bray93},
and Liu and Mazenko 
\cite{Mazenko92}, that the OJK approximation becomes exact as the number
 of dimensions approaches infinity.
  Evaluating the dimensional dependence of the 
first order correction term enables this hypothesis to be tested.

The main result of this paper is that the first order correction
 to the correlation function is $O(1/d^2)$, 
lending weight to the assertion that the OJK theory becomes exact
 in an infinite
 number of dimensions.
  The exact form of the leading order correction to
the two-time correlation function, $C_1({\bf r}, t_1, t_2)$,
 is calculated in the large-$d$ limit; this is found to obey both
 Porod's Law \cite{Porod}, and the Tomita sum rule 
\cite{Tomita84,Tomita86}, with the singular
contribution in the correction term modifying the amplitude of the
Porod tail, and in the limit $t_1\gg t_2$, it
 is found that the correction term is of exactly the same form as the
 zero order result.
  The complexity of the calculation, however, has so far 
 prevented the evaluation of higher order terms. 

In the following section we present an outline of the OJK calculation,
which provides the starting point for the perturbation calculation, we
then in section \ref{sec:Perturb} proceed with a detailed description of the
perturbation calculation.  We conclude with a summary and discussion
of the results.

\section{The OJK Theory }
\label{tech:OJKderive}

Consider a system described by a non-conserved scalar order
parameter. 
The evolution of this system following a rapid quench from a high
temperature homogeneous phase to a regime where there are two
equilibrium phases is
described by the TDGL equation,
\begin{equation}
{\partial\phi({\bf x},t)\over\partial t}=
-\lambda{\delta F[\phi]\over\delta\phi},
\label{intro:TDGL}
\end{equation}
where $F[\phi] $ is a Ginzburg-Landau free energy functional, given by
\begin{equation}
 F[\phi]=\int d^d{\bf x}\left(
 {1\over 2}|\mbox{\boldmath $\nabla$}\phi|^2
+V(\phi)  \right).
\label{intro:energy}
\end{equation}
 $V(\phi) $ is a potential whose minima define the equilibrium values
of the order parameter; the
conventional choice in the case of a scalar order parameter is 
$V(\phi) = (1-\phi^2)^2/4$.
Note that there is no noise term in this equation, hence the results
are only valid for quenches to temperatures where 
the effect of thermal fluctuations is negligible.

 Following the work of Ohta, Jasnow and Kawasaki \cite{OJK},
 the scalar field,
$\phi({\bf r},t)$, is replaced by a smoothly varying auxiliary field,
$m({\bf r},t)$, where $\phi({\bf r},t)$ and $m({\bf r},t)$ have the
same sign.  The zeros of the field $m$ then define the positions of
the domain walls, and the normal, $\hat{\bf n}$, to a domain wall in the
direction of increasing $\phi$ is given by
\begin{equation}
\hat{\bf n}= {\mbox{\boldmath $\nabla$}m\over |\mbox{\boldmath $\nabla$}m|}.
\end{equation} 
Inserting this into the Allen-Cahn equation for the velocity of an
interface \cite{Allen}, $v=-K= -\mbox{\boldmath $\nabla$}.\hat{\bf n}$;
$K$ is the curvature of the interface, we obtain:
\begin{equation}
v={1\over |\mbox{\boldmath $\nabla$}m |}\left(
-\mbox{\boldmath $\nabla$}^2 m+ n_in_j\,
{\partial^2\,m\over \partial x_i\partial x_j}
\right),
\label{techojk:vel1}
\end{equation}
where $v$ is the speed of the interface in the direction of $\hat{\bf n}$.
We now seek to obtain an evolution equation for $m({\bf r},t)$ by
 linking the domain-wall velocity to the time-dependence of
the auxiliary field.
The total time derivative of the auxiliary field in a frame of
reference moving at a velocity ${\bf v}$ is given by
\begin{equation}
{dm\over dt}={\partial m\over \partial t}+{\bf v}.\mbox{\boldmath $\nabla$}m.
\end{equation}
If this frame is moving with the interface velocity, then  
${\bf v}=v\hat{\bf n}$ and 
$\mbox{\boldmath $\nabla$}m $ are parallel, and the total
derivative of $m$ vanishes, implying that
\begin{equation}
{\partial m({\bf r},t)\over \partial t}=- v|\mbox{\boldmath
$\nabla$}m|.
\label{tech:mad}
\end{equation}
We now substitute the expression for the interface velocity
(from equation (\ref{techojk:vel1})) into equation (\ref{tech:mad}) to obtain,
\begin{equation}
{\partial m\over\partial t} = \mbox{\boldmath $\nabla$}^2m -n_in_j
{\partial^2 m\over\partial x_i \partial x_j}.
\label{techojk:evol}
\end{equation}
 This is the time evolution equation for the auxiliary field obtained
by OJK.

To make analytic progress, we now linearise this partial differential
equation by replacing  $n_in_j$  by its spherical
average $\delta_{ij}/d$.  In this approximation, the evolution of the field is 
governed by a simple diffusion equation with diffusion constant
$D=1-1/d$.
To calculate the correlation function we need to express the original
order parameter $\phi$ in terms of the auxiliary field.  In the thin
wall limit this is given by $\phi = \hbox{sgn} (m)$, therefore
\begin{equation}
  C({\bf r},t_1,t_2) = \langle\hbox{sgn}\,(m({\bf x}+{\bf r},t_1))
  \hbox{sgn} (m({\bf x},t_2))\rangle.
\end{equation}
 It is convenient to choose Gaussian initial conditions
for the auxiliary field, with zero mean, and correlator
\begin{equation}
\langle m({\bf x}+{\bf r},0)\, m({\bf x},0)\rangle=\Delta\,\delta ({\bf r}).
\label{techojk:initial}
\end{equation}
Since the evolution equation is linear, the auxiliary field will then have
a Gaussian distribution at all times.  Therefore to calculate the
correlation function we simply need to evaluate the joint probability
distribution for $m({\bf x}+{\bf r},t_1)\equiv m(1) $ and 
$m({\bf x},t_2)\equiv m(2) $.
This is given by 
\begin{eqnarray}
&&P(m(1),m(2))=
{1\over 2\pi(S(1)S(2)(1-\gamma^2))^{1/2}}
\nonumber\\[0.3cm]
&&\qquad\qquad\times\exp\left[-{1\over 2(1-\gamma^2)}\left(
{m(1)^2\over S(1)}+{m(2)^2\over S(2) }
-2\gamma{m(1)m(2)\over (S(1)S(2))^{1/2} }
\right)
\right],
\end{eqnarray}
where $\gamma$ is the normalized correlator,
\begin{eqnarray}
\gamma&=&{\langle\,m(1)\,m(2)\,\rangle\over
\langle\,m(1)^2\,\rangle^{1/2}
\langle\,m(2)^2\,\rangle^{1/2}}
\\&=&
\left({4t_1t_2\over (t_1+t_2)^2}\right)^{d/4}\exp\left[-{{\bf
r}^2\over 4D(t_1+t_2)}\right],
\label{techojk:gamma}
\end{eqnarray}
$ S(1)=\,\langle\,m(1)^2\,\rangle$ and $S(2)=
\,\langle\,m(2)^2\,\rangle
$ \cite{Review}.
Completion of the average over the field $m$ yields the final OJK
result for the correlation function, and is given by
\begin{equation}
C({\bf r},t_1,t_2)={2\over\pi}\sin^{-1}\gamma.
\label{techojk:exact}
\end{equation}
We note that this result only has a trivial dependence on the
dimension $d$, through the diffusion constant, $D=1-1/d$.

\section{Perturbation Theory}       
\label{sec:Perturb}

The starting point for this calculation is the OJK equation for the
evolution of the auxiliary field $m({\bf x},t)$ (derived in section 
\ref{tech:OJKderive}),
\begin{equation}
{\partial m\over\partial t} = \mbox{\boldmath $\nabla$}^2m -n_in_j
{\partial^2 m\over\partial x_i \partial x_j},
\label{ojk:eq1}
\end{equation}
where ${\bf n}=\mbox{\boldmath $\nabla$}m/|\mbox{\boldmath $\nabla$}m|$.
As noted in the previous section, this equation is highly non-linear 
and to make analytic progress OJK 
 approximated the non-linear term,
$n_in_j$, by its spherical average, $\delta_{ij}/d$, reducing
equation (\ref{ojk:eq1}) 
  to the diffusion equation with diffusion
constant $D=1-1/d$.

 In this calculation, the terms dropped in this
 approximation are treated as a perturbation to the diffusion equation
and the first order correction to the correlation function is
calculated.  Therefore the equation we wish to solve is 
\begin{equation}
{\partial m\over \partial t} = D\mbox{\boldmath $\nabla$}^2m
-\lambda\left(n_in_j-{\delta_{ij}\over d}\right)
{\partial^2 m\over\partial x_i \partial x_j}  ,
\label{ojk:ojk1}
\end{equation}
where $\lambda$ is a small perturbation parameter.

The solution of equation (\ref{ojk:ojk1}) can be expressed as a power
 series in $\lambda$, $m({\bf r},t)=m_0+\lambda m_1+
O(\lambda^2)$; substituting this back into equation (\ref{ojk:ojk1})
and equating orders of $\lambda$ gives two coupled
partial differential equations for $m_0$ and $m_1$, the solutions to
 which are given by
\begin{eqnarray}
m_0({\bf r},t)&=&\int d^d\mu\, \,
G({\bf r}-\mbox{\boldmath $\mu$},t)\, m_0(\mbox{\boldmath $\mu$},0),
\label{ojk:m0}
\\
m_1({\bf r},t)&=&-\int_0^t d\tau \int d^d\mu \,\,
G({\bf r}-\mbox{\boldmath $\mu$},t-\tau) \nonumber\\
&&\qquad\qquad\times \left(
{\partial_i m_0(\mbox{\boldmath $\mu$},\tau)\partial_j
 m_0(\mbox{\boldmath $\mu$},\tau)\over
|\mbox{\boldmath $\nabla$}m_0(\mbox{\boldmath $\mu$},\tau)|^2}-
{\delta_{ij}\over d} \right)
{\partial^2 m_0(\mbox{\boldmath $\mu$},\tau)\over
\partial\mu_i \partial\mu_j } ,
\label{ojk:m1}
\end{eqnarray}
where $G({\bf \mu}-{\bf r},\tau-t)$ is the Green's function for the
diffusion equation and is given by
\begin{equation}
G({\bf r},t)={1 \over (4\pi Dt)^{d/2}} \exp\left[-{{\bf r}^2\over 4Dt}\right],
\label{ojk:green}
\end{equation}
and $\partial_i m_0(\mbox{\boldmath $\mu$},\tau)\equiv 
{\partial m_0(\mbox{\boldmath $\mu$},\tau)\over \partial\mu_i}$.

In the thin wall limit the order parameter is related to the auxiliary
 field by the equation $\phi=\hbox{sgn} (m)$.  Hence, by using the 
 integral representation of $\hbox{sgn} (m)$ and expanding the
 expression for the correlation function in $\lambda$, we find that
\begin{equation}
  C({\bf r},t_1,t_2)= \,\langle\phi ({\bf x}+{\bf r},t_1)\,\phi ({\bf x},t_2)\rangle\,
  = C_0({\bf r},t_1,t_2)+\lambda C_1({\bf r},t_1,t_2) + O(\lambda^2),
\end{equation}
where
\begin{eqnarray}
  C_0({\bf r},t_1,t_2)& = &\langle\hbox{sgn}\,(m_0({\bf x}+{\bf r},t_1))
  \hbox{sgn} (m_0({\bf x},t_2))\rangle, \\ 
 C_1({\bf r},t_1,t_2)& = &2\biggl(
\langle\hbox{sgn}(m_0({\bf x}+{\bf r},t_1)) \,\delta(m_0({\bf x},t_2))\,
m_1({\bf x},t_2) \rangle \nonumber \\ 
&&\quad+ \langle\hbox{sgn}(m_0({\bf x},t_2)) \,\delta(m_0({\bf x}+{\bf r},t_1))\,
m_1({\bf x}+{\bf r},t_1)\rangle \biggr).
\end{eqnarray}

If we define $\tilde{C}_1({\bf r},t_1,t_2)$ by
\begin{equation}
\tilde{C}_1({\bf r},t_1,t_2)=2
\langle\hbox{sgn}(m_0({\bf x}+{\bf r},t_1)) \,\delta(m_0({\bf x},t_2))\,
m_1({\bf x},t_2) \rangle,
\label{ojk:c1}
\end{equation}
then the first order correction to the correlation function is given
by
\begin{equation}
C_1({\bf r},t_1,t_2)=\tilde{C}_1({\bf r},t_1,t_2)+
\tilde{C}_1(-{\bf r},t_2,t_1).
\label{ojk:correlation}
\end{equation}

Since the two terms on the RHS of the expression for $C_1({\bf
 r},t_1,t_2)$ (equation (\ref{ojk:correlation})) differ only in
 that $t_1 \rightarrow t_2$ and ${\bf r} \rightarrow -{\bf r}$,
 we will only deal
 with one of these averaged terms, $\tilde{C}_1({\bf r},t_1,t_2)$,
 evaluating the
complete expression at the end of the calculation.

  Substituting for $m_1$ from equation (\ref{ojk:m1}) into equation 
(\ref{ojk:c1}), we obtain the following expression for 
 $\tilde{C}_1({\bf r},t_1,t_2)$,
\begin{eqnarray}
&&\tilde{C}_1({\bf r},t_1,t_2)=-2\int_0^{t_2} d\tau \int d^d\mu \,\,
G({\bf x}-\mbox{\boldmath $\mu$},t_2-\tau)\langle
\,\hbox{sgn}[m_0({\bf x}+{\bf r},t_1)]
\nonumber\\ 
&&\qquad\times
\left(
{\partial_i m_0(\mbox{\boldmath $\mu$}
,\tau)\partial_j m_0(\mbox{\boldmath $\mu$}
,\tau)\over
|\mbox{\boldmath $\nabla$}m_0(\mbox{\boldmath $\mu$}
,\tau)|^2}-{\delta_{ij}\over d} \right)
\delta(m_0({\bf x},t_2))
{\partial^2 m_0(\mbox{\boldmath $\mu$},\tau)\over
\partial\mu_i \partial\mu_j }\,\rangle.
\label{ojk:eq2}
\end{eqnarray}
We now impose the conventional Gaussian initial conditions, with a zero 
mean, and a correlator given by
\begin{equation}
\langle m({\bf x}+{\bf r},0)\, m({\bf x},0)\rangle=\Delta\,\delta ({\bf r}),
\end{equation}
The average on the RHS of equation (\ref{ojk:eq2}) can be
evaluated if we extract the differential operator ${\partial^2\over
\partial \mu_i \partial \mu_i}$ from the average by defining a new spatial
variable $\mbox{\boldmath $\nu$} $. Equation (\ref{ojk:eq2}) can then
be written as
\begin{equation}
\tilde{C}_1({\bf r},t_1,t_2)=-2\int_0^{t_2} d\tau \int d^d\mu \,\,
G({\bf x}-\mbox{\boldmath $\mu$}
,t_2-\tau) 
\left.{\partial^2 I_{ij}\over \partial\nu_i \partial\nu_j} 
\right |_{\mbox{\boldmath $\nu$}=\mbox{\boldmath $\mu$}},
\label{ojk:start}
\end{equation}
where
\begin{equation}
I_{ij}=
\langle\left(
{\partial_i m_0(\mbox{\boldmath $\mu$}
,\tau)\partial_j m_0(\mbox{\boldmath $\mu$}
,\tau)\over
|\mbox{\boldmath $\nabla$}m_0(\mbox{\boldmath $\mu$}
,\tau)|^2}-{\delta_{ij}\over d} \right)
\,\hbox{sgn}(m_0({\bf x}+{\bf r},t_1))\,m_0(\mbox{\boldmath $\nu$},\tau)
\,\delta(m_0({\bf x},t_2))\,
\rangle.
\label{ojk:istart}
\end{equation}
The ensemble average in equation (\ref{ojk:istart}) can now be
completed by using the joint probability 
distribution connecting the variables  
$m_0({\bf x}+{\bf r},t_1)$, $m_0({\bf x},t_2)$,
$m_0(\mbox{\boldmath $\nu$},\tau)$ and  
${\partial m_0(\mbox{\boldmath $\mu$},\tau)\over\partial \mu_i}$.

For simplicity, at this point we introduce a contracted notation:
\begin{eqnarray}
&{\partial m_0(\mbox{\boldmath $\mu$},\tau)\over\partial \mu_i}  \to
m'_i,
\qquad\qquad &
m_0(\mbox{\boldmath $\nu$},\tau)  \to m(2),  \\ 
&m_0({\bf x}+{\bf r},t_1)  \to m(1), \qquad\qquad &
m_0({\bf x},t_2)  \to  m(3),
\label{ojk:notation}
\end{eqnarray}
and define the  vector $\tilde{{\bf m}}$  by
\begin{equation}
\tilde{{\bf m}}=
\left (m'_1\,,m'_2\,,m'_3\,, \cdots\,,m'_d\,,m(1)\,,m(2)\,,m(3)\,\right ).
\label{ojk:m}
\end{equation}
The joint probability distribution for the components of this vector
is then given by,
\begin{equation}
P(\tilde{{\bf m}})=
{1\over (2\pi)^{d+3\over 2}(\hbox{det}A^{-1})^{1\over 2}}
\,\exp(-{1\over 2}\tilde{m}_i\,A_{ij}\,\tilde{m}_j), 
\end{equation}
where  $A^{-1}_{ij}=\langle\tilde{m}_i\tilde{m}_j\rangle$.
$P(\tilde{\bf m})$ is evaluated in appendix \ref{ojk:App1}
and is given by
\begin{equation}
 P(\tilde{\bf m})={1\over (2\pi)^{d+3\over 2}(\hbox{det}A^{-1})^{1\over 2}}
 \,\exp\left[-{F(\tilde{\bf m})\over 2\hbox{det}A^{-1}} \right] ,
\label{ojk:prob}
\end{equation}
where
\begin{eqnarray}
F(\tilde{\bf m})&=&
 m'_k\theta_{kl}m'_l +2\mbox{\boldmath $\xi $}.{\bf m'}m(1) +
  2\mbox{\boldmath $\eta $}.{\bf m'}m(2)
 + 2\mbox{\boldmath $\zeta$}.{\bf m'}m(3) +pm(1)^2+qm(2)^2\nonumber\\
 &&\qquad+rm(3)^2+2sm(1)m(2)+2tm(1)m(3)+2um(2)m(3)
 .
\label{ojk:f1}
\end{eqnarray}
and the expressions for the coefficients $\mbox{\boldmath $\xi $}$, 
$\mbox{\boldmath $\eta $}$, $\mbox{\boldmath $\zeta $}$ and   
$p$, $q$, $\dots$ $u$ are given in appendix \ref{ojk:App1}.

Equation (\ref{ojk:istart}) may therefore be written in the form  
\begin{eqnarray}
I_{ij}
&=&\int  d^d {\bf m'}\int_{-\infty}^{\infty} dm(1)
\int_{-\infty}^{\infty} dm(2)\int_{-\infty}^{\infty} dm(3)\nonumber \\  
&&\qquad \times P(\tilde{\bf m})
\left({m'_i\,m'_j\over|{\bf m'}|^2}-{\delta_{ij}\over d} \right)
\,\hbox{sgn}[m(1)]\,m(2)\,\delta[m(3)].
\label{ojk:average1}
\end{eqnarray}

The function $\tilde{C}_1({\bf r},t_1,t_2)$ defined by equation 
(\ref{ojk:start}) is evaluated in three main steps.  Firstly, the
expression for $I_{ij}$ (equation (\ref{ojk:average1})) is evaluated.
 This result is then differentiated to obtain 
$\left.{\partial^2 I_{ij}\over \partial\nu_i \partial\nu_j} 
\right |_{\mbox{\boldmath $\nu$}=\mbox{\boldmath $\mu$}}$, and
finally, 
this expression is substituted back into equation (\ref{ojk:start}) and the
remaining integrals are completed.

To evaluate equation (\ref{ojk:average1}), we must complete the integrals over
$\tilde{\bf m}$.
The integral over $m(3)$ is trivial; on setting $m(3)$ to zero in
equation (\ref{ojk:average1}), we see that the integral over
$m(2)$ can be reduced to a Gaussian by completing the
square in the exponent of the probability density function, 
$-F(\tilde{\bf m})/ 2\hbox{det}A^{-1}$.
  For simplicity we 
deal with the expression for $F(\tilde{\bf m})$ (equation
(\ref{ojk:f1})) first.  Completing the square using the 
substitution $m'(2)=m(2)+(sm(1)+\mbox{\boldmath $\eta$}.{\bf m'})/q $ 
in $m(2)$ gives
\begin{equation}
F(\tilde{m})=m'_k\left(\theta_{kl}-{\eta_k\eta_l\over q}\right)m'_l
+{(pq-s^2)\over q}\,m(1)^2
-2{(s\mbox{\boldmath $\eta$}-q\mbox{\boldmath $\xi$})\over q}.
{\bf  m'}\,m(1) + qm'(2)^2.
\label{ojk:f21}
\end{equation}
This may be simplified further by substituting for $\mbox{\boldmath $\xi$}$
 and $\mbox{\boldmath $\eta$}$ in the factor
 $s\mbox{\boldmath $\eta$}-q\mbox{\boldmath $\xi$}$,
 from equations (\ref{ojk:xi}) and
 (\ref{ojk:eta}) respectively, to give
\begin{equation}
s\mbox{\boldmath $\eta$}-q\mbox{\boldmath $\xi$}
=(pq-s^2){\bf a}-(us-qt){\bf c}=(pq-s^2){\bf h},
\label{ojk:seat}
\end{equation}
where this equation is used to define the vector ${\bf h}$. 
Using equations (\ref{ojk:pq-ss}) and (\ref{ojk:us-qt}), ${\bf h}$ may
then be written as
\begin{equation}
{\bf h}= {\bf a}
 - \left(us-qt\over pq-s^2\right) {\bf c}=
{\bf a}
 - \left(zy-{\bf a.c}\over z\lambda_3-{\bf c}^2\right) {\bf c}.
\label{ojk:h}
\end{equation}
On substituting equation (\ref{ojk:seat}) back into equation
(\ref{ojk:f21}), $F(\tilde{\bf m})$  reduces to
\begin{equation}
F(\tilde{m})=m'_k\left(\theta_{kl}-{\eta_k\eta_l\over q}\right)m'_l
+{(pq-s^2)\over q}\,\biggl(m(1)^2-2{\bf h}.{\bf m}'\,m(1)\biggr)
 + qm'(2)^2.
\label{ojk:f13}
\end{equation}
Substituting equations (\ref{ojk:prob}) and (\ref{ojk:f13}) back into equation
(\ref{ojk:average1}) gives
\begin{eqnarray}
\nonumber
 I_{ij}&=&{(2\pi)^{-(d+3)/2}\over (\hbox{det}A^{-1})^{1/2}}
\int  d^d {\bf m'} \,\left({m'_i\,m'_j\over|{\bf m'}|^2}-
             {\delta_{ij}\over d}\right)
\exp\left[-{1\over 2\hbox{det}A^{-1}}m'_k(\theta_{kl}-
                 {\eta_k\eta_l\over q})m'_l \right] \\ 
\nonumber
&&\times\int_{-\infty}^{\infty} dm(1) \,\hbox{sgn}[m(1)]
\exp\left[-{(pq-s^2)\over 2q\,
\hbox{det}A^{-1}}\,(m(1)^2-2{\bf h}.{\bf m'}m(1))\right] \nonumber\\ 
&&\times\int_{-\infty}^{\infty} dm'(2) \,\left(
m'(2)-{(sm(1)+\mbox{\boldmath $\eta$}.{\bf m'})\over q}  \right)
\exp\left[-{qm'(2)^2\over 2\hbox{det}A^{-1}}\right].
\end{eqnarray}
The $m(2)$ integral is now in the form of a
Gaussian, on completion of which, $I_{ij}$ simplifies to
\begin{eqnarray}
\nonumber
I_{ij}&=&{-(2\pi)^{-(d+2)/2}\over q^{3/2}}
\int  d^d {\bf m'} \,\left({m'_i\,m'_j\over|{\bf m'}|^2}-
             {\delta_{ij}\over d}\right)
\exp\left[-{1\over 2\hbox{det}A^{-1}}m'_k(\theta_{kl}-
                 {\eta_k\eta_l\over q})m'_l \right] \\ 
&&\times\int_{-\infty}^{\infty} dm(1) \,\hbox{sgn}[m(1)]
(sm(1)+\mbox{\boldmath $\eta$}.{\bf m'})\nonumber \\ 
&&\qquad\times\exp\left[-{(pq-s^2)\over 2q\,
\hbox{det}A^{-1}}\,(m(1)^2-2{\bf h}.{\bf m'}m(1))\right] .
\label{ojk:Im2}
\end{eqnarray}

We now manipulate this expression into a
form in which the ${\bf m'}$ integral can be completed.
Firstly we need to remove the $|{\bf m'}|^2$ term from the
denominator; this can be achieved by rewriting the denominator
as an integral over an exponential,
\begin{equation}
{1\over |{\bf m'}|^2}=
\int_0^{\infty}d\tilde{v}\exp\left[-\tilde{v}|{\bf m'}|^2 \right].
\label{ojk:denom}
\end{equation} 
The second step is to manipulate the $m(1)$ dependence so
that the exponent in the complete expression for $I_{ij}$ can be
written in the form ${\bf m'}^T\Omega{\bf m'}$, where $\Omega$ is a $d\times
d$ matrix to be determined later.

Due to the presence of the $\hbox{sgn}[m(1)]$ function, the positive
and negative ranges of the integral must be treated separately; in each
case we complete the square in the exponent using the substitution
$\tilde{u}=(m(1)-{\bf h}.{\bf m'})/{\bf h}.{\bf m'}$.
The anti-symmetric contribution to the $m(1)$ integral in equation
(\ref{ojk:Im2}) may be evaluated
directly, and the symmetric contribution reduces to a term proportional
to an error function.
This integral, therefore, reduces to
\begin{eqnarray}
&&\int dm(1) \,\hbox{sgn}[m(1)]
(sm(1)+\mbox{\boldmath $\eta$}.{\bf m'})
\exp\left[-{(pq-s^2)\over 2q\,
\hbox{det}A^{-1}}\,(m(1)^2-2{\bf h}.{\bf m'}m(1))\right]
\nonumber\\
&& =\alpha+\beta_{kl}m'_km'_l\int_0^1d\tilde{u}\exp\left[- 
{(pq-s^2)\over 2q\,\hbox{det}A^{-1}}({\bf h}.{\bf m'})^2(\tilde{u}^2-1)
  \right],
\label{ojk:intm1}
\end{eqnarray}
where
\begin{equation}
\alpha = {2qs\, \hbox{det}A^{-1} \over (pq-s^2)} \qquad\hbox{and}\qquad 
\beta_{kl}  = 2h_k(\eta_l+sh_l).
\label{ab}
\end{equation}
Substituting equations (\ref{ojk:denom}) and (\ref{ojk:intm1}) back
into the expression for 
$I_{ij}$ (equation (\ref{ojk:Im2})), we obtain
\begin{eqnarray}
I_{ij}&=&{1\over (2\pi)^{(d+2)/2}q^{3/2}}
\int   d^d {\bf m'}\,
\exp\left[-{1\over 2\hbox{det}A^{-1}}m'_e(\theta_{ef}-
                 {\eta_e\eta_f\over q})m'_f \right]\nonumber\\
&&\times\left( {\delta_{ij}\over d}-
m'_i\,m'_j\int_0^{\infty}d\tilde{v}
\exp\left[-\tilde{v}|{\bf m'}|^2 \right] \right) \nonumber \\
&&\times\left(\alpha+\beta_{kl}\,m_k'm_l'\int_0^1d\tilde{u}\exp\left[- 
{(pq-s^2)\over 2q\,\hbox{det}A^{-1}}({\bf h}.{\bf m'})^2(\tilde{u}^2-1)
  \right] \right).
\label{ojk:exi}
\end{eqnarray} 

We are now in a position to define the general matrix
$\Omega(\tilde{u},\tilde{v})$ referred to earlier;
\begin{equation}
\Omega_{ij}(\tilde{u},\tilde{v})={1\over \hbox{det}A^{-1}}
\left(\theta_{ij}-{\eta_i\eta_j\over q}
+{(pq-s^2)\over q}h_ih_j(\tilde{u}^2-1)\right)+2\tilde{v}\delta_{ij}.
\label{ojk:omega}
\end{equation}
  By expanding the
RHS of equation (\ref{ojk:exi}), we see that the exponent of each
contribution is given by  the general $d\times
d$ matrix $\Omega(\tilde{u},\tilde{v})$, evaluated at different values
of $\tilde{u}$ and $\tilde{v}$.
Equation (\ref{ojk:exi}) can therefore be written as the sum of four
standard integrals,
\begin{equation}
I_{ij}=\sum_{n=1}^4\,I_n^{ij},
\end{equation}
where
\begin{eqnarray}
I_1^{ij}&=&{\alpha\,\delta_{ij}\over d(2\pi)^{(d+2)/2}q^{3/2}}
\int  d^d {\bf m'} \,
exp\left[-{1\over 2}{\bf m'}^T\Omega(1,0){\bf m'}\right]\nonumber
\\ 
&=&{\alpha\delta_{ij}\over 2\pi d q^{3/2}(\hbox{det}\,\Omega(1,0))^{1/2}} 
\label{ojk:I1} ,\\ 
I_2^{ij}&=&-{\alpha\over (2\pi)^{(d+2)/2}q^{3/2}}
\int_0^{\infty}d\tilde{v}\,\int  d^d {\bf m'} \,
m'_im'_j\exp\left[-{1\over 2} {\bf m'}^T\Omega(1,\tilde{v}){\bf
m'}\right]\nonumber \\ 
& =&-{\alpha\over 2\pi q^{3/2}}
\int_0^{\infty}d\tilde{v}\,
  {\Omega^{-1}_{ij}(1,\tilde{v})\over(\hbox{det}\Omega(1,\tilde{v}))^{1/2}}
                       , \label{ojk:I2} \\
I_3^{ij}&=&{\beta_{kl}\,\delta_{ij}\over d(2\pi)^{(d+2)/2}q^{3/2}}
\int_0^1d\tilde{u}\int d^d {\bf m'} \, 
m'_km'_l\exp\left[-{1\over 2} {\bf m'}^T\Omega(\tilde{u},0){\bf
  m'}\right] \nonumber \\ 
&=&{\delta_{ij}\,\beta_{kl}\over 2\pi dq^{3/2}}
\int_0^1d\tilde{u}{\Omega^{-1}_{kl}(\tilde{u},0)  \over
 (\hbox{det}\,\Omega(\tilde{u},0))^{1/2}}
 \label{ojk:I3}, \\
I_4^{ij}&=&-{\beta_{kl}\over (2\pi)^{(d+2)/2}q^{3/2}}
\int_0^{\infty}d\tilde{v}\int_0^1d\tilde{u}\,\int d^d {\bf m'} \,
m'_im'_jm'_km'_l\exp\left[-{1\over 2}{\bf
  m'}^T\Omega(\tilde{u},\tilde{v}){\bf m'}\right] \nonumber \\
&=&-{\beta_{kl}\over 2\pi q^{3/2}} 
\int_0^{\infty}d\tilde{v}\,\int_0^1d\tilde{u}
{1\over (\hbox{det}\,\Omega(\tilde{u},\tilde{v}))^{1/2}} 
\nonumber \\
&&
\times\left(
\Omega^{-1}_{ij}(\tilde{u},\tilde{v})
  \Omega^{-1}_{kl}(\tilde{u},\tilde{v})
  +\Omega^{-1}_{ik}(\tilde{u},\tilde{v})
  \Omega^{-1}_{jl}(\tilde{u},\tilde{v}) 
+\Omega^{-1}_{il}(\tilde{u},\tilde{v})\Omega^{-1}_{jk}(\tilde{u},\tilde{v})
 \right). \label{ojk:I4}
\end{eqnarray}
Details of the calculation of the inverse and determinant of $\Omega(\tilde{u},\tilde{v})$
are supplied in appendix \ref{ojk:App2}.  They are given by:
\begin{eqnarray}
\Omega^{-1}_{ij}(\tilde{u},\tilde{v})&=&{\lambda\over \Lambda z}\left(
\delta_{ij}-{c_ic_j\over (\Lambda z\lambda_3-(\Lambda -1){\bf c}^2)}
-{\tilde{u}^2\,k_ik_j\over \tilde{q}\Delta 
(\Lambda z\lambda_3-(\Lambda -1){\bf c}^2)}
\right),\qquad
\label{ojk:inv} \\
\hbox{det}\,\Omega(\tilde{u},\tilde{v})
&=&\Lambda^{d-2}\left({z\over\lambda}\right)^d\,\Delta(\tilde{u},\tilde{v}),
\label{ojk:det}
\end{eqnarray} 
where 
 \begin{eqnarray}
\Delta(\tilde{u},\tilde{v} )&=&
\Lambda\left({\Lambda z\lambda_3-(\Lambda -1){\bf c}^2\over
z\lambda_3-{\bf c}^2}\right)-{\Lambda z\lambda_3\tilde{u}^2\over
(z\lambda_3-{\bf c}^2)}+{\Lambda z^2\tilde{u}^2\over\tilde{q}}
(\lambda_1\lambda_3-y^2) \nonumber \\
&&\quad-{(\Lambda -1) \tilde{u}^2\over\tilde{q}}
({\bf a}^2{\bf c}^2-{\bf a}.{\bf c}^2),
\label{ojk:delta} \\
k_i(\tilde{v})&=&(\Lambda z\lambda_3-(\Lambda -1){\bf c}^2)\,a_i-
(\Lambda zy-(\Lambda -1){\bf a}.{\bf c})\,c_j,
\label{ojk:k} \\
\hbox{and}\qquad\qquad\tilde{q}&=&\left({z\over\lambda}\right)^{d+2}q=
 (z\lambda_1 -{\bf a}^2 )(z\lambda_3 -{\bf c}^2 )- (zy-{\bf a}.{\bf
 c})^2. 
\label{tildeq}
\end{eqnarray}

At this point we can calculate 
$\left.{\partial^2 I_{ij}\over \partial\nu_i \partial\nu_j} 
\right |_{\mbox{\boldmath $\nu$}=\mbox{\boldmath $\mu$}} $, which once
completed will be substituted into equation (\ref{ojk:start}). 
 Firstly, we notice that the only variables which depend
 on $\mbox{\boldmath$\nu$}$ in the expressions for the $I_n^{ij}$ are 
$\alpha$ and $\beta_{kl}$ (defined in (\ref{ab})).
Our first step therefore, is to
simplify these quantities and then calculate their derivatives.

Inserting equation (\ref{ojk:pq-ss}) into the expression for $\alpha$ and
differentiating, we find
\begin{equation}
\left.{\partial^2\,\alpha\over \partial\nu_i \partial\nu_j}
\right|_{\mbox{\boldmath $\mu$}
}=
{2q\over (z\lambda_3-{\bf c}^2)}
\left({z\over\lambda}\right)^{d+1}
\left.{\partial^2\,s\over \partial\nu_i \partial\nu_j}
\right|_{\mbox{\boldmath $\mu$}
},
\label{ojk:dalpha}
\end{equation}
(From this point on  we will use the notation 
$\left.{\partial^2\,\alpha\over \partial\nu_i \partial\nu_j}
\right|_{\mbox{\boldmath $\mu$}}$ to indicate that we are evaluating the
differential at $\mbox{\boldmath $\nu$}=\mbox{\boldmath $\mu$} $.)
To simplify  $\beta_{kl}$, we substitute for 
 $\bf h $  and $\mbox{\boldmath $\eta$}$
 from equations (\ref{ojk:h}) and (\ref{ojk:eta}) respectively; 
 using equation (\ref{ojk:squ}) we find that
\begin{eqnarray}
\beta_{kl}  = {2q\over z\lambda_3-{\bf c}^2}
\Biggl[a_k-\left({zy-a.c\over z\lambda_3-{\bf c}^2}\right)c_k \Biggr]
\Biggl[(zv-{\bf b.c})\,c_l-(z\lambda_3-{\bf c}^2)\,b_l \Biggr].
\end{eqnarray}
We now note that, from the definition of ${\bf b}$ (equation
(\ref{ojk:a})),
\begin{equation}
\left.{\partial^2\,{\bf b}\over \partial\nu_i \partial\nu_j}
\right|_{\mbox{\boldmath $\mu$}
}=0     ,
\label{ojk:db}
\end{equation}
so the derivative of $\beta_{kl}$ is given by
\begin{equation}
\left.{\partial^2\,\beta_{kl}\over \partial\nu_i \partial\nu_j}
\right|_{\mbox{\boldmath $\mu$}
}={2qz\over (z\lambda_3-{\bf c}^2)}
\left.{\partial^2\,v\over \partial\nu_i \partial\nu_j}
\right|_{\mbox{\boldmath $\mu$}
}\,h_kc_l.
\label{ojk:dbeta}
\end{equation}

Differentiating equations (\ref{ojk:I1})-(\ref{ojk:I4}) and
using equations (\ref{ojk:det}), (\ref{ojk:dalpha}) and
(\ref{ojk:dbeta}), we obtain the following expressions for the
$\left.{\partial^2\,I_n^{ij}\over \partial\nu_i \partial\nu_j}
\right|_{\mbox{\boldmath $\mu$}} $: 
\begin{eqnarray}
\left.{\partial^2\,I_1^{ij}\over \partial\nu_i \partial\nu_j}
\right|_{\mbox{\boldmath $\mu$}
}&=&{\delta_{ij}\over d\pi 
(z\lambda_3-{\bf c}^2)[\tilde{q}\Delta(1,0)]^{1/2}}
\left({z\over\lambda}\right)^{d+2} 
\left.{\partial^2\,s\over \partial\nu_i \partial\nu_j}
\right|_{\mbox{\boldmath $\mu$}
}\label{ojk:dI1}, \\
\left.{\partial^2\,I_2^{ij}\over \partial\nu_i \partial\nu_j}
\right|_{\mbox{\boldmath $\mu$}
}&=&
{-1\over \pi (z\lambda_3-{\bf c}^2)}
\left({z\over\lambda}\right)^{d+2} 
\left.{\partial^2\,s\over \partial\nu_i \partial\nu_j}
\right|_{\mbox{\boldmath $\mu$}
}
\int_0^{\infty}{d\tilde{v}\,\Omega^{-1}_{ij}(1,\tilde{v})\over
\Lambda^{(d-2)/2}[\tilde{q}\Delta(1,\tilde{v})]^{1/2}}\label{ojk:dI2}, \\
\left.{\partial^2\,I_3^{ij}\over \partial\nu_i \partial\nu_j}
\right|_{\mbox{\boldmath $\mu$}
}&=&
{z^2\,\delta_{ij}h_kc_l\over
d \pi\lambda (z\lambda_3-{\bf c}^2)}
\left.{\partial^2\,v\over \partial\nu_i \partial\nu_j}
\right|_{\mbox{\boldmath $\mu$}
}
\int_0^1 d\tilde{u}{\Omega^{-1}_{kl}(\tilde{u},0)\over
[\tilde{q}\Delta(\tilde{u},0)]^{1/2}}
\label{ojk:dI3},
\\
\left.{\partial^2\,I_4^{ij}\over \partial\nu_i \partial\nu_j}
\right|_{\mbox{\boldmath $\mu$}
}&=&{-z^2\,h_kc_l\over
\pi\lambda (z\lambda_3-{\bf c}^2)}
\left.{\partial^2\,v\over \partial\nu_i \partial\nu_j}
\right|_{\mbox{\boldmath $\mu$}
}\,\int_0^{\infty}d\tilde{v}\,\int_0^1d\tilde{u}
{1\over\Lambda^{(d-2)/2}[\tilde{q}\Delta(\tilde{u},\tilde{v})]^{1/2} } 
\nonumber\\ &&\times
\biggl(
\Omega^{-1}_{ij}(\tilde{u},\tilde{v})
  \Omega^{-1}_{kl}(\tilde{u},\tilde{v})
 +\Omega^{-1}_{ik}(\tilde{u},\tilde{v})
  \Omega^{-1}_{jl}(\tilde{u},\tilde{v}) \nonumber \\ &&\qquad\qquad
+\Omega^{-1}_{il}(\tilde{u},\tilde{v})
\Omega^{-1}_{jk}(\tilde{u},\tilde{v})\biggr)
\label{ojk:dI4}.
\end{eqnarray}

At this point we shall pause to give an overview of the simplification
of equations (\ref{ojk:dI1})-(\ref{ojk:dI4}). We first consider 
equations (\ref{ojk:dI3}) and (\ref{ojk:dI4}); on
completing the contraction over the $k$ and $l$ indices, we obtain
expressions in which the $\tilde{u}$ integration may be completed
exactly.  We then complete the contractions over the  $i$ and $j$ indices
in the expression
\begin{equation}
\left.{\partial^2\,I_{ij}\over \partial\nu_i \partial\nu_j}
\right|_{\mbox{\boldmath $\mu$}}
=\sum_{n=1}^4\,
\left.{\partial^2\,I_n^{ij}\over \partial\nu_i \partial\nu_j}
\right|_{\mbox{\boldmath $\mu$}},
\label{ojk:sum}
\end{equation}
and substitute this  back into the
expression for $\tilde{C}_1({\bf r},t_1,t_2)$ (equation
\ref{ojk:start}).

Before we complete the contraction over the
 $k$ and $l$ indices in equations (\ref{ojk:dI3}) and (\ref{ojk:dI4}),
 we will derive some results which 
will be used later. 
 Using the definitions of ${\bf h}$, $\Delta$, ${\bf k}$ and $\tilde{q}$ 
(equations  (\ref{ojk:h}), (\ref{ojk:delta}), (\ref{ojk:k}) and (\ref{tildeq})
 respectively), we find 
\begin{equation}
{\bf k}(\tilde{v}).{\bf h}=
{\tilde{q}\over\tilde{u}^2}\left(
\Delta(\tilde{u},\tilde{v})
 - \Lambda \left({\Lambda z\lambda_3-(\Lambda -1){\bf c}^2\over
z\lambda_3-{\bf c}^2}\right)\right).
\label{ojk:kdoth}
\end{equation}
Using equation (\ref{ojk:kdoth})
 together with the definitions of 
$\Omega^{-1}(\tilde{u},\tilde{v})$, ${\bf h}$ and ${\bf k}$ (equations
(\ref{ojk:inv}), (\ref{ojk:h}) and (\ref{ojk:k}) respectively) we also 
 obtain:
\begin{eqnarray}
\Omega^{-1}_{ij}(\tilde{u},\tilde{v})h_j&=&
{\lambda k_i(\tilde{v})
\over z(z\lambda_3-{\bf c}^2)\Delta(\tilde{u},\tilde{v})}, 
\label{ojk:omega-h} \\ 
\Omega^{-1}_{ij}(\tilde{u},\tilde{v})c_j&=&
{\lambda\over z(\Lambda z\lambda_3-(\Lambda - 1){\bf c}^2)}\left(
(z\lambda_3-{\bf c}^2)c_j - {{\bf k}(\tilde{v}).{\bf c}\,\tilde{u}^2\over
\Lambda\tilde{q}\Delta(\tilde{u},\tilde{v})}k_j(\tilde{v})\right).
\label{ojk:omega-c}
\end{eqnarray} 
Substituting equation (\ref{ojk:omega-h}) and (\ref{ojk:omega-c})
 into equations (\ref{ojk:dI3}) and (\ref{ojk:dI4}),
we can therefore complete the contractions over the $k$ and
 $l$ indices to obtain:
\begin{eqnarray}
\left.{\partial^2\,I_3^{ij}\over \partial\nu_i \partial\nu_j}
 \right|_{\mbox{\boldmath $\mu$}
}&=&{z\,{\bf k}(0).{\bf c}\,\delta_{ij}\over
d\pi \tilde{q}^{1/2}(z\lambda_3-{\bf c}^2)^2}
\left.{\partial^2\,v\over \partial\nu_i \partial\nu_j}
\right|_{\mbox{\boldmath $\mu$}
}
\int_0^1{d\tilde{u}\over\Delta(\tilde{u},0)^{3/2}},
\label{ojk:ddI3}\\
\left.{\partial^2\,I_4^{ij}\over \partial\nu_i \partial\nu_j}
 \right|_{\mbox{\boldmath $\mu$}
}&=&
-{z\over
\pi \tilde{q}^{1/2}(z\lambda_3-{\bf c}^2)^2 }
\left.{\partial^2\,v\over \partial\nu_i \partial\nu_j}
\right|_{\mbox{\boldmath $\mu$}
}\,
\int_0^{\infty}\int_0^1{d\tilde{u}d\tilde{v}\over\Lambda^{(d-2)/2}
\Delta(\tilde{u},\tilde{v})^{3/2}}
\,\nonumber\\
&& \times \Biggl[
{\lambda{\bf k}(\tilde{v}).{\bf c}\over\Lambda z}
\left(\delta_{ij} -{c_ic_j\over \Lambda z\lambda_3-(\Lambda - 1){\bf
 c}^2} -{3k_i(\tilde{v})k_j(\tilde{v})\,\tilde{u}^2\over
 \tilde{q}\Delta(\tilde{u},\tilde{v})
(\Lambda z\lambda_3-(\Lambda - 1){\bf c}^2)} 
 \right)\nonumber\\
&&\qquad+{\lambda(z\lambda_3-{\bf c}^2)(c_ik_j(\tilde{v})+c_jk_i(\tilde{v}))
\over z (\Lambda z\lambda_3-(\Lambda - 1){\bf
 c}^2)} 
\Biggr].
\label{ojk:ddI4}
 \end{eqnarray}
Equations (\ref{ojk:ddI3}) and (\ref{ojk:ddI4}) are now in a form in
which we can  complete the 
$\tilde{u}$ integration.   
 We  see that the only integrals required are 
$\int_0^1\,d\tilde{u}\,\Delta(\tilde{u},\tilde{v})^{-3/2}$ and 
$\int_0^1\,d\tilde{u}\, \tilde{u}^2\Delta(\tilde{u},\tilde{v})^{-5/2}$; since
 $\Delta (\tilde{u},\tilde{v})$ (which is 
defined by equation (\ref{ojk:delta}))
 is of the form $\gamma + \delta\tilde{u}^2$, these integrals
 over $\tilde{u}$ may easily be completed to give:
\begin{eqnarray}
\int_0^1d\tilde{u}{1\over [\Delta(\tilde{u},\tilde{v})]^{3/2}} &=& 
{(z\lambda_3-{\bf c}^2)\over
\Lambda (\Lambda z\lambda_3-(\Lambda -1){\bf c}^2)
[\Delta(1,\tilde{v})]^{1/2}},\label{ojk:dint1}\\ 
\int_0^1d\tilde{u}{\tilde{u}^2\over [\Delta(\tilde{u},\tilde{v})]^{5/2}} &=& 
{(z\lambda_3-{\bf c}^2)\over
3\Lambda (\Lambda z\lambda_3-(\Lambda -1){\bf c}^2)
[\Delta(1,\tilde{v})]^{3/2}},\label{ojk:dint2}
\end{eqnarray}   
where 
\begin{eqnarray}
\tilde{q}\Delta (1,\tilde{v})&=&
\Lambda^2z^2(\lambda_1\lambda_3-y^2 )-\Lambda(\Lambda -1)z
(\lambda_3{\bf a^2}-2y{\bf a.c}+\lambda_1{\bf c}^2)
\nonumber \\ &&\qquad
+(\Lambda -1)^2({\bf a}^2{\bf c}^2-{\bf a}.{\bf c}^2).
\label{ojk:delta1}
\end{eqnarray}
We can therefore 
substitute equations (\ref{ojk:dint1}) and (\ref{ojk:dint2}) into
 the RHS of equations (\ref{ojk:ddI3})
and (\ref{ojk:ddI4}) to obtain:
\begin{eqnarray}
\left.{\partial^2\,I_3^{ij}\over \partial\nu_i \partial\nu_j}
\right|_{\mbox{\boldmath $\mu$}
}
&=& {{\bf k}(0).{\bf c}\,\delta_{ij}\over d \pi \lambda_3
 (z\lambda_3-{\bf c}^2)
[\tilde{q}\Delta(1,0)]^{1/2}}
\left.{\partial^2\,v\over \partial\nu_i \partial\nu_j}
\right|_{\mbox{\boldmath $\mu$}
}
\label{ojk:g2},
\\
\left.{\partial^2\,I_4^{ij}\over \partial\nu_i \partial\nu_j}
\right|_{\mbox{\boldmath $\mu$}
}
&=&-{z\over\pi (z\lambda_3-{\bf c}^2)}
\left.{\partial^2\,v\over \partial\nu_i \partial\nu_j}
\right|_{\mbox{\boldmath $\mu$}
}
\int_0^{\infty}{d\tilde{v}\over\Lambda^{d/2}
[\tilde{q}\Delta(1,\tilde{v})]^{1/2}}
\,\nonumber\\ 
\qquad\qquad&&\times\left(
{{\bf k}(\tilde{v}).{\bf c}\Omega^{-1}_{ij}(1,\tilde{v})\over
(\Lambda z\lambda_3-(\Lambda -1){\bf c}^2)}+
{\lambda (z\lambda_3 - {\bf c}^2)(k_i(\tilde{v})c_j+c_ik_j(\tilde{v}))\over
z\,(\Lambda z\lambda_3-(\Lambda -1){\bf c}^2)^2 }
\right) 
\label{ojk:g1}.\qquad
\end{eqnarray}

We now  substitute equations (\ref{ojk:dI1}), (\ref{ojk:dI2}),
 (\ref{ojk:g2}) and  (\ref{ojk:g1})  into equation 
(\ref{ojk:sum}) and complete the contraction over the $i$ and
$j$ indices.  The details of this calculation are contained within
 appendix \ref{ojk:App3}.  
Following the contraction over the  $i$ and
$j$ indices, equation (\ref{ojk:sum}) reduces to
\begin{equation}
\left.{\partial^2\,I_{ij}\over \partial\nu_i \partial\nu_j}
\right|_{\mbox{\boldmath $\mu$}}
=-{\lambda\over\pi z}\int_0^{\infty}d\tilde{v}\sum_{n=1}^6T_n,
\label{ojk:diffI}
\end{equation}
where the terms $T_n$ are defined by the equations 
(\ref{ojk:t1})-(\ref{ojk:t6}).
Substituting equations (\ref{ojk:green}) and (\ref{ojk:diffI}) into
equation (\ref{ojk:start}),
where without loss of generality we can set ${\bf x}=0$, we obtain the
following expression for $\tilde{C}_1({\bf r},t_1,t_2)$:
\begin{equation}
\tilde{C}_1({\bf r},t_1,t_2)=
\int_0^{t_2}d\tau\int_0^{\infty}d\tilde{v}\int d^d\mu\,
{2\lambda\sum_{n=1}^6  T_n  \over \pi z(4\pi D(t_2-\tau))^{d/2} }
 \exp\left(-{\mu^2 \over 4D(t_2-\tau)}\right).
\label{tilc}
\end{equation}

The spatial integral over ${\bf\mu}$ may be completed by transforming to 
spherical polars and choosing a change of variables which allows the integral
 to be completed by steepest descents.  In spherical polars the spatial
integral may be written as 
\begin{equation}
\int d^d\mu \,= {2\,(\pi)^{d-1\over 2}\over\Gamma({d-1\over 2})}
\int_0^{\infty} d\mu\,\mu^{d-1}\,
\int_0^{\pi}d\psi\,(\sin\psi )^{d-2} .
\end{equation}
We notice  that each expression for $T_n$ (equations 
(\ref{ojk:t1})-(\ref{ojk:t6})) contains the factor
$\Lambda^{-d/2}=(1+2\lambda\tilde{v}/z)^{-d/2}$; we therefore make the
substitution $\tilde{w}=\tilde{v}\lambda d/z$, so that in the
large-$d$ limit, this factor reduces to an exponential in $\tilde{w}$.
We also make the additional substitution $\tau=t_2(1-x/d)$, and
rescale $\mu$ to 
 $\hat{\mu}=(xt_2)^{-1/2}\mu$. The form of the $\tau$ substitution is
 chosen in the expectation that the contribution to the integral from later
 times will dominate the result. 
Equation (\ref{tilc}) then reduces to 
\begin{eqnarray}
\tilde{C}_1({\bf r},t_1,t_2) &=&{4t_2\over\pi^{3/2} d^2\Gamma ({d-1\over2})}
\left({d\over 4D}\right)^{d/2}\sum_{n=1}^6\nonumber \\
&&\times
\int_0^{d}dx\int_0^{\infty}d\tilde{w}
\int_0^\infty {d\hat{\mu}\over\hat{\mu}}\int_0^\pi{d\psi\over (\sin\psi)^2}
\exp[-d\,g(\hat{\mu},\psi)]\,
\,T_n,
\label{ojk:corrsub}
\end{eqnarray}
where
\begin{equation}
g(\hat{\mu},\psi)={\hat{\mu}^2\over 4D}-\ln (\hat{\mu}\sin\psi).
\end{equation}

Once the integral is in this form, the $\hat{\mu}$ and $\psi$
integrals can be completed in the large-$d$ limit by steepest
descents.  Applying this limit does not
represent a serious limitation on the calculation, since this is the 
 regime of most interest, as explained
in the introduction.
In the limit $d\to\infty$, provided the large-$d$ behavior of the terms
$T_i$ is controlled (as demonstrated in the first
section of appendix \ref{ojk:App4}), the value of the integral is dominated by
the contribution from the neighbourhood around the minima of the function
$g(\hat{\mu},\psi)$. 
 
Within the range of integration, the function
$g(\hat{\mu},\psi)$  has a global minimum at
 $\hat{\mu}=\sqrt{2D}$, $\psi =\pi/2$. We can expand the integrand to
first order about this minimum, reducing the $\hat{\mu}$ and
$\psi$ integrations
in equation (\ref{ojk:corrsub}) into Gaussians, and giving
\begin{eqnarray}
&&\tilde{C}_1({\bf r},t_1,t_2)={4t_2\exp(-d/2)
\over (2D)^{1/2}\pi^{3/2}d^2\Gamma({d-1\over 2}) }
\left({d\over 2}\right)^{d/2}\sum_{n=1}^6\int_0^d
dx\,\int_0^{\infty}d\tilde{w}
\left.T_n
\right|_{\mu=\sqrt{2D},\psi =\pi /2} \nonumber\\
&&\qquad\times\int_0^{\infty}d\hat{\mu}\,
\exp\left({-d (\hat{\mu}-\sqrt{2D})^2\over 2D}\right)
\int_0^{\pi}d\psi\,
\exp\left({-d(\psi-\pi/2)^2  \over 2}\right).
\label{ojk:cc}
\end{eqnarray}
Since we are only interested in the first order 
correction to the correlation function, we use Stirling's formula to
 evaluate the leading order contribution from the Gamma function,
which is given by:
\begin{equation}
\Gamma\left({d-1\over 2}\right)\sim 
{2(2\pi)^{1/2}\over d}\left({d\over 2}\right)^{d/2}\exp(-d/2).
\label{ojk:gam}
\end{equation}
On substituting equation (\ref{ojk:gam}) back into equation (\ref{ojk:cc}) and
completing the Gaussian integrals, we find
\begin{equation}
\tilde{C}_1({\bf r},t_1,t_2)={2t_2\over\pi d^2}
\int_0^\infty\!\!\!\int_0^\infty\!\!\!dx\,d\tilde{w}\,\,
\left.\sum_{n=1}^6T_n
\right|_{\mu=\sqrt{2D},\psi=\pi /2}.
\label{ojk:correx}
\end{equation}

Now all that remains is to calculate the leading order
contribution from the integration over $x$ and
$\tilde{w}$ of the terms $T_n$, evaluated at 
$\mu=\sqrt{2D}$ and $ \psi =\pi /2$.
  Each of the terms may be expanded as a
power-series in $1/d$, and in section 2 of appendix \ref{ojk:App4}
each term is evaluated to $O(1)$, giving:

\begin{eqnarray}
T_1&=&{\gamma(t_2-t_1)\exp(-x/2)\,\tilde{w}\exp(-\tilde{w})\over
8Dt_2(t_1+t_2)(1-\gamma^2)^{3/2}}
\left(x\left(1-{4t_1t_2\gamma^2\over(t_1+t_2)^2}\right)
+{2t_2\gamma^2{\bf r}^2\over D(t_1+t_2)^2}
\right),\label{red1}\quad\quad \\ [0.2cm]
T_2&=&{\gamma (t_2-t_1)x\exp(-x/2)\,\tilde{w}\exp(-\tilde{w})\over
4 Dt_2(t_1+t_2)(1-\gamma^2)^{1/2}},  \\[0.2cm]
T_3&=&{\gamma (t_2-t_1)x(x-2)\exp(-x/2)\,\exp(-\tilde{w})\over 
8 Dt_2(t_1+t_2)(1-\gamma^2)^{1/2}}, \\[0.2cm]
T_4&=&{-\gamma{\bf r}^2x\exp(-x/2)\exp(-\tilde{w})\over 
8 D^2(t_1+t_2)^2(1-\gamma^2)^{1/2}},  \\[0.2cm]
T_5&=&{-\gamma\exp(-x/2)\exp(-\tilde{w})\over 8Dt_2(1-\gamma^2)^{3/2}}
\left(x\left(1-{4t_1t_2\gamma^2\over(t_1+t_2)^2}\right)
+{2t_2\gamma^2{\bf r}^2\over D(t_1+t_2)^2}
\right)\nonumber \\
&&\qquad\times \left[ {t_2-t_1\over t_1+t_2} -{t_2{\bf r}^2\over D(t_1+t_2)^2}
+{x\over 2}\left(1-{4t_2^2\over (t_1+t_2)^2}\right)\right], \\[0.2cm]
T_6&=&0.  \label{red2}
\end{eqnarray}

We now complete the $x$ and $\tilde{w}$ integrations of the
expressions given in equations (\ref{red1})-(\ref{red2}), finding:
\begin{eqnarray}
\int_0^\infty\!\!\!\int_0^\infty\!\!\!dx\,d\tilde{w}\,\,
T_1 &=& {\gamma(t_2-t_1)\over
2Dt_2(t_1+t_2)(1-\gamma^2)^{3/2}}
\left(1-{4t_1t_2\gamma^2\over(t_1+t_2)^2}
+{t_2\gamma^2{\bf r}^2\over D(t_1+t_2)^2}
\right),\quad\quad \label{ojk:ttt1}\\ [0.2cm]
\int_0^\infty\!\!\!\int_0^\infty\!\!\!dx\,d\tilde{w}\,\,
T_2 &=& {\gamma (t_2-t_1)\over Dt_2(t_1+t_2)(1-\gamma^2)^{1/2}},  \\[0.2cm]
\int_0^\infty\!\!\!\int_0^\infty\!\!\!dx\,d\tilde{w}\,\,
T_3 &=& {\gamma (t_2-t_1)\over 
Dt_2(t_1+t_2)(1-\gamma^2)^{1/2}}, \\[0.2cm]
\int_0^\infty\!\!\!\int_0^\infty\!\!\!dx\,d\tilde{w}\,\,
T_4 &=& {-\gamma{\bf r}^2\over 
2 D^2(t_1+t_2)^2(1-\gamma^2)^{1/2}}, \\[0.2cm]
\int_0^\infty\!\!\!\int_0^\infty\!\!\!dx\,d\tilde{w}\,\,
T_5 &=& {-\gamma\over 2Dt_2(1-\gamma^2)^{3/2}}
\Biggl[\left(1-{4t_2^2\over (t_1+t_2)^2}\right)
\left(2-{8t_1t_2\gamma^2\over(t_1+t_2)^2}+{t_2\gamma^2{\bf r}^2\over D(t_1+t_2)^2} \right)
\nonumber \\&&
\!\!\!\!
+ \left( {t_2-t_1\over t_1+t_2} -
{t_2{\bf r}^2\over D(t_1+t_2)^2} \right)
\left(1-{4t_1t_2\gamma^2\over(t_1+t_2)^2}+{t_2\gamma^2{\bf r}^2\over D(t_1+t_2)^2} \right)
\Biggr],
 \\[0.2cm]
\int_0^\infty\!\!\!\int_0^\infty\!\!\!dx\,d\tilde{w}\,\,
T_6 &=& 0 . \label{ojk:ttt6}
\end{eqnarray}

On substituting equations (\ref{ojk:ttt1})-(\ref{ojk:ttt6}) into equation
 (\ref{ojk:correx}), we obtain: 
\begin{eqnarray}
\tilde{C}_1({\bf r},t_1,t_2)&=& {2\gamma\over D\pi d^2(1-\gamma^2)^{3/2}}
\times\Biggl[
\left(1-{4t_1t_2\gamma^2\over (t_1+t_2)^2}\right)
\left({4t_2^2\over (t_1+t_2)^2}-1\right) \nonumber
\\
&&+{2(t_2-t_1)(1-\gamma^2)\over (t_1+t_2)}
+{2t_2^2(t_2-t_1)\gamma^2{\bf r}^2\over
D(t_1+t_2)^4 }+
{t_2^2\gamma^2{\bf r}^4\over
2D^2(t_1+t_2)^4 }
\Biggr].
\label{ojk:resultS}
\end{eqnarray}
Finally, on substituting equation (\ref{ojk:resultS})  back into
 equation (\ref{ojk:correlation}), we obtain the complete expression
 for the first order correction to the correlation function:
\begin{eqnarray}
&&C_1({\bf r},t_1,t_2)=  {4\gamma
\over D\pi d^2 (1-\gamma^2)^{3/2}}\nonumber \\
&&\quad\times\Biggl[
\left({t_1-t_2\over t_1+t_2}\right)^2
\left(1-{4t_1t_2\gamma^2\over (t_1+t_2)^2}
+{\gamma^2{\bf r}^2\over D(t_1+t_2)} \right)  
+{(t_1^2+t_2^2)\gamma^2{\bf r}^4\over 4D^2(t_1+t_2)^4}
\Biggr]. \label{ojk:RESULT}
\end{eqnarray}

 We now express our final result for the first order correction to the
correlation function in the same form as the exact OJK result
(equation (\ref{techojk:exact})).  In this form
 the correlation function is given by
\begin{equation}
C({\bf r},t_1,t_2)={2\over\pi}\sin^{-1}[\gamma(1+\lambda F({\bf r},t_1,t_2))],
\end{equation}
where
\begin{eqnarray}
&&F({\bf r},t_1,t_2)={2\over D d^2(1-\gamma^2)}
\nonumber \\ &&\qquad\times\Biggl[
{(t_1-t_2)^2\over (t_1+t_2)^2}
\left(1-{4t_1t_2\gamma^2\over (t_1+t_2)^2}+
{\gamma^2{\bf r}^2\over D(t_1+t_2)}
\right) 
+{(t_1^2+t_2^2)\gamma^2{\bf r}^4\over 4D^2(t_1+t_2)^4}
\Biggr].
\end{eqnarray}

We can clearly see from this expression that the first order
correction to the correlation function is $O(1/d^2)$, which
lends weight to the assertion that the OJK result becomes exact in
an infinite-dimensional system.

\subsection{Two special cases}

We now evaluate this result in two special cases: at
equal times, and when the times are widely separated, 
$t_1\gg t_2$.

\begin{figure}[htbp]
\centerline{\epsfxsize=7.0cm \epsfbox{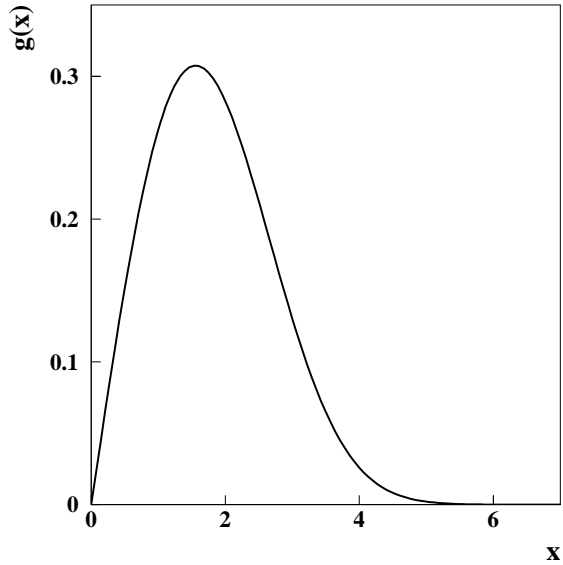}}
\caption{Scaling function for the first order correction to the
correlation function.}
\label{ojkc1}
\end{figure}

At equal times, the first order correction term is given by 
\begin{equation}
C_1({\bf r},t)={1\over 8D\pi d^2}g\left({r\over(Dt)^{1/2}} \right),
\label{ojk:pert}
\end{equation}
where
\begin{equation}
g(x)={x^4\exp(-3x^2/8)\over (1-\exp(-x^2/4))^{3/2}}.
\end{equation}
This correction term clearly exhibits the expected scaling, $L\sim
t^{1/2}$; the scaling function $g(x)$ is shown in figure
\ref{ojkc1}.

\vspace{-0.3cm}

\begin{figure}[htbp]
\centerline{\epsfxsize=7.0cm \epsfbox{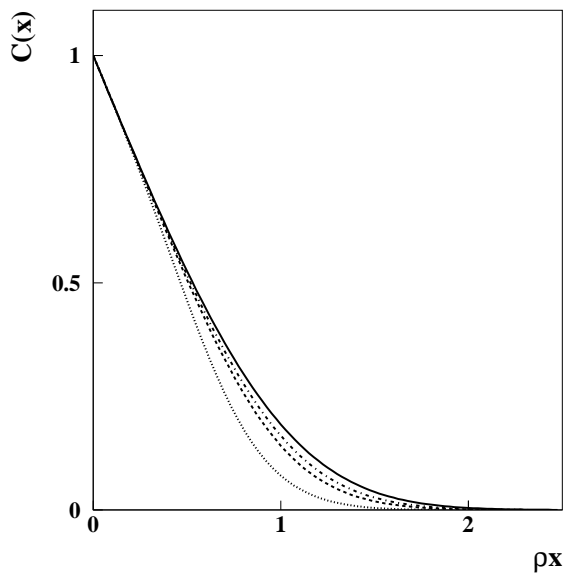}}
\caption{Comparison of the OJK and perturbed correlation functions,
shown for $d=2,3,4$. The solid line is the OJK result, which is
independent of dimension.  As the dimension increases, the
size of the perturbation decreases.}
\label{ojkcompare}
\end{figure}

Figure \ref{ojkcompare} shows a comparison of the zero order OJK
result (equation (\ref{techojk:exact})) with the perturbed result for $d=2,3$ and
$4$ at equal times;  the functions have all been scaled so that
 they have the same gradient at the origin.
Although the result is only valid for large $d$, 
the figure clearly demonstrates
that the perturbation will have the effect of lowering the exact OJK
result.  This is discussed further in section \ref{ojkdiscus}.
\vspace{-0.3cm}

If we now expand equation (\ref{ojk:pert}) about $r=0$, we see
that for small $r$,
\begin{equation}
C_1({\bf r},t)\sim {1\over D\pi d^2} {r\over (Dt)^{1/2}}
\left(1-{3r^2\over 16Dt}+ {7r^4\over 512D^2t^2}+O(r^6/(Dt)^3) \right)
\label{ojk:expresult}
\end{equation}
and hence the result obeys Porod's Law \cite{Porod} and the Tomita sum
rule \cite{Tomita84,Tomita86}. 

We now consider the case where the times are widely separated.  If
 $t_1 \gg t_2$, then equation (\ref{ojk:RESULT}) reduces to 
\begin{equation}
C_1({\bf r},t_1, t_2)={4\over D\pi d^2}\left({4t_2\over t_1}
\right)^{d/4}
\exp\left(-{{\bf r}^2\over 4Dt_1} \right).
\label{okj:one}
\end{equation}
Comparing this to the conventional scaling form,
\begin{equation}
C_1({\bf r},t_1, t_2)\sim \left({L_2\over
L_1}\right)^{\overline{\lambda}}\,h\left({r\over
L_1}\right)\qquad\qquad t_1\gg t_2,
\end{equation}
where  $L_1$ and $L_2$ are the characteristic lengths for
 the system at times $t_1$ and $t_2$ respectively, we find that 
$\overline{\lambda}=d/2$, as expected.
We also note that the leading order correction to the correlation
 function in this limit has exactly the same form  as the zero order
 correlation function; from equation (\ref{techojk:exact}) we have
\begin{equation}
C_0({\bf r},t_1, t_2)={2\over \pi }\left({4t_2\over t_1}
\right)^{d/4}
\exp\left(-{{\bf r}^2\over 4Dt_1} \right).
\label{okj:zero}
\end{equation}

\section{Discussion}
\label{ojkdiscus}

In this paper I have described a perturbation theory applicable to
OJK theory in a system with an infinite number of dimensions and no
noise.  The leading order
correction term to the two-time correlation function has been
calculated by treating the
non-linear terms in the OJK auxiliary-field evolution equation as a
perturbation to the linearized equation.  There are two separate
 motivations for this study.
Firstly, since in conventional OJK theory, the evolution equation is
linear and Gaussian initial conditions are imposed, it follows
 that the distribution of the auxiliary field must be Gaussian at
all times.    This assumption, present in
many of the approximate theories, has been critically assessed by 
Yeung {\it et. al.} \cite{Yeung94}; they show that the auxiliary
field distribution measured directly
from numerical simulations is not exactly Gaussian.  Hence,
 the
advantage of the approach outlined above 
is that the auxiliary field distribution
may take any form.

  Secondly, it has been proposed by several 
authors \cite{Bray93,Mazenko92} that OJK theory is exact in an
infinite number of dimensions.  Determining the
dimensional dependence of the correction term  allows
this hypothesis to be examined.

The main result of this paper is that the first order correction
term to the correlation function is $O(1/d^2)$, which gives further
confidence to the assertion that OJK theory becomes exact in the
large-$d$ limit.  This may be compared to the calculation of  Liu and
  Mazenko \cite{Mazenko92} in which they made a perturbative
 expansion near $d=1$ and $d=\infty$ about the approximate theory
developed by Mazenko \cite{Mazenko89,Mazenko90,Mazenko91}.
  At large $d$, the first order correction term in
their expansion was found to be  O(1/d). 
The relative sizes of these corrections 
are consistent with numerical simulations \cite{Humayun92c} which
show  that the OJK result provides a marginally more accurate
prediction of simulation data for the pair correlation function
 than the more
sophisticated approach of Mazenko.  However both  results do
suggest that the results of OJK are asymptotically exact in 
infinite-dimensional systems.
We also note that
in a recent work,
 Mazenko \cite{Mazenko97} has developed a perturbation
expansion in the cumulants, which at zeroth order recovers the OJK
result. At second order, however, significant
deviations from the OJK theory are obtained at large $d$, in contrast to the
$O(1/d^2)$ correction obtained here.

The two-time correction term to the correlation function is given by
equation (\ref{ojk:RESULT}).  In the limit $t_1\gg t_2$ (on comparing
equations (\ref{okj:one}) and (\ref{okj:zero})), we find that the correction term
has exactly the same form as the zero order result, and  the expected OJK
result, $\overline{\lambda}=d/2$, is recovered.

If we examine the leading order correction at equal times,
 given by equation (\ref{ojk:pert}), the first
observation is that this function is odd in $r$, as is the
zero order term. Porod's law \cite{Porod} is therefore obeyed;
  the $O(r)$ term in the expansion (equation (\ref{ojk:expresult}))
modifies the amplitude of the large-$k$ tail, which is proportional to the
density of defects \cite{Review,Porod,Humayun93a}.
  The absence of an $r^2$ term ensures that 
the correction term satisfies the Tomita sum rule 
\cite{Tomita84,Tomita86},  $\int_0^\infty\,dk(k^{d+1}S(k)-A)=0$,
where $A$ is the amplitude of the Porod tail.
 
We also note that the
perturbation has a negligible effect on the large-distance behavior of the
correlation function, see figure \ref{ojkc1}.

Figure \ref{ojkcompare} demonstrates how the perturbation 
 modifies the OJK
result, plotting $C_0+C_1$ for $d=2,3,4$.  Although this calculation is only valid at
large d, this graph demonstrates that the correction term will have the
effect of lowering the OJK result.  This is exactly as expected since
 Humayun and Bray \cite{Humayun92c} showed by comparing 
simulation results with OJK theory, that while the OJK result  
provides an accurate prediction for
initial conditions with short-range correlations,
 the theoretical result is slightly higher than the simulation data.
 However, it has also been demonstrated that when long-range 
correlation are present in the
initial conditions, the OJK results are no longer satisfactory \cite{Humayun92c}.
 This suggests that a possible extension to this work could be to
consider the effects of long-range initial conditions on the 
calculation.   

The main limitation of this calculation is that to retrieve the full
evolution equation for the auxiliary field (equation (\ref{techojk:evol})), 
we need to set the perturbation parameter, $\lambda$ to $1$. This means that
 the calculation of
the entire correction term requires a sum over all orders in
$\lambda$.  However, due to the complexity of the
present calculation, I have been unable to evaluate the correction
terms at higher orders in $\lambda$.
  In principle the sum of the higher order terms
could alter the $d$-dependence of the correction term, but the
result remains a strong indication that OJK theory becomes exact in
an infinite-dimensional system.

Finally, since we noted that the distribution of the auxiliary field is
non-Gaussian, it is of interest to consider the exact form of this
distribution. This can be calculated as follows; the distribution can
be written in the following form, $P(x)=\langle\delta(x-m({\bf
r},t))\rangle$, this may be expanded in $\lambda$ using $m({\bf r},t)=
m_0({\bf r},t)+\lambda m_1({\bf r},t) + O(\lambda^2)$, to give
$P(x)=\langle\delta(x-m_0({\bf r},t))\rangle-\lambda
\langle m_1({\bf r},t){d\,\over dx}\delta(x-m_0({\bf r},t))\rangle$.
The first term on the RHS reduces to the expected Gaussian
distribution;  the first order term in $\lambda$ can be evaluated
using a similar method to the correlation function calculation,
i.e. by inserting the expression for $m_1({\bf r},t)$, multiplying by
the relevant probability distribution function, and completing the
integrals.  However we will leave this question to future work.

%+++++++++++++++++++++++++++++++++++++++++++++++++++++++++++
%+++++++++++++++++++++++++++++++++++++++++++++++++++++++++++

\section{Acknowledgements}
This work was  supported by EPSRC (United Kingdom).
  I would like to thank Tim Newman, Sarah Phillipson
 and especially Alan Bray for many useful discussions.

\appendix

%+++++++++++++++++++++++++++++++++++++++++++++++++++++++++++
%+++++++++++++++++++++++++++++++++++++++++++++++++++++++++++

\section{Evaluation of the Joint Probability Distribution}
\label{ojk:App1}
\indent

In this appendix the joint probability distribution is explicitly
calculated.  It is defined by
\begin{equation}
P(\tilde{\bf{m}})=
{1\over (2\pi)^{d+3\over 2}(\hbox{det}A^{-1})^{1\over 2}}
\,\exp\left[-{1\over 2}\tilde{m}_i\,A_{ij}\,\tilde{m}_j\right],
\label{app1:pp} 
\end{equation}
where 
$A^{-1}_{ij}=\langle\tilde{m}_i\tilde{m}_j\rangle $ and the vector ${\bf
  \tilde{m}}$ is defined in equation (\ref{ojk:m}).  The first step is to
calculate all the correlators which define the elements of the matrix
 $A^{-1}$; we can then find the inverse of this matrix ($A_{ij}$) and
 the determinant, $\hbox{det}\,A^{-1}$.

\subsection{Calculation of the Correlators}

In this section we evaluate all the elements of the matrix 
 $A^{-1}_{ij}=\langle\tilde{m}_i\tilde{m}_j\rangle $; for this we
require the following correlators:
$\langle m_0({\bf x},t)\,m_0({\bf x}',t')\rangle$, 
$\langle m_0({\bf x},t)\,{\partial\over\partial x'_i}m_0({\bf
 x}',t')\rangle$, and
$\langle {\partial\over\partial x_i}m_0({\bf x},t)\,{\partial\over\partial
    x_j}m_0({\bf x},t)\rangle$.

Substituting equation (\ref{ojk:green}) into equation (\ref{ojk:m0})
gives an expression for the auxiliary field $m_0({\bf x},t)$:
\begin{equation}
m_0({\bf x},t)
={1\over (4\pi Dt)^{d\over 2}}\int d^2\mu\,\,
m_0(\mbox{\boldmath $\mu$},0)
\exp \left[-{({\bf x}-\mbox{\boldmath $\mu$})^2\over 4Dt}
\right],
\end{equation}
and substituting this expression into the correlator 
$\langle m_0({\bf x},t)\,m_0({\bf x}',t')\rangle$ gives
\begin{eqnarray}
 && \langle m_0({\bf x},t)\,m_0({\bf x}',t')\rangle ={1\over (4\pi D)^d(tt')^{d\over
      2}}
\nonumber \\
&&\qquad\times\int d^d\mu\int d^d\eta\,\, \Delta\delta(\mbox{\boldmath
    $\mu$}-\mbox{\boldmath $\eta$}) \exp\left[ -{({\bf
      x}-\mbox{\boldmath $\mu$})^2\over 4Dt} -{({\bf
      x'}-\mbox{\boldmath $\eta$})^2\over 4Dt'} \right],
\end{eqnarray}
where we have already applied the conventional Gaussian initial conditions,

\noindent
$\langle m_0({\bf x}+{\bf r},0)\,m_0({\bf x},0)\rangle 
=\Delta\delta({\bf r})$.

This integral may be evaluated by completing the square in the
exponent and making a change of variables $\mbox{\boldmath
  $\mu$}'=\mbox{\boldmath $\mu$}-(t'{\bf x}+t{\bf x}')/(t+t')$,
leaving a simple Gaussian integral which, once completed, gives
\begin{equation}
\langle m_0({\bf x},t)\,m_0({\bf x}',t')\rangle\,  
= {\Delta\over (4\pi D(t+t'))^{d/2}}\,
\exp\left[ -{({\bf x}-{\bf x'})^2\over 4D(t+t')}\right].
\label{app1:corr1}
\end{equation}

The remaining correlators are easily obtained by differentiating equation
(\ref{app1:corr1}); we obtain:
\begin{eqnarray}
\langle m_0({\bf x},t)\,{\partial\over\partial x'_i}m_0({\bf x}',t')\rangle &=&
 {(x_i-x'_i)\over 2D(t+t')}\,\langle m_0({\bf x},t)\,m_0({\bf
 x}',t')\rangle  ,\label{app1:corr2}\\
  \langle {\partial\over\partial x_i}m_0({\bf x},t)\,{\partial\over\partial
    x_j}m_0({\bf x},t)\rangle &=& {\delta_{ij}\over 4Dt}
\langle m_0({\bf  x},t)\,m_0({\bf x},t)\rangle  .\label{app1:corr3}
\end{eqnarray}

Substituting equations (\ref{app1:corr1}), (\ref{app1:corr2}) 
and (\ref{app1:corr3}) into the definition of $A^{-1}$

\noindent
($A^{-1}_{ij}=\langle\tilde{m}_i\tilde{m}_j\rangle $), we find
\begin{equation}
A^{-1}=({\lambda\over z})\pmatrix{
 1         &         & 0       & \vdots      &\vdots       & \vdots  \cr
           & \ddots  &         &{\bf a}&{\bf b}&{\bf c} \cr
 0         &         & 1       & \vdots      & \vdots      & \vdots  \cr
\cdots&{\bf a}^T& \cdots  &z\lambda_1   & zw         & zy      \cr
\cdots&{\bf b}^T& \cdots  & zw          & z\lambda_2 & zv      \cr
\cdots&{\bf c}^T& \cdots  & zy          &  zv         &z\lambda_3 \cr
},
\end{equation}
where: 
\begin{eqnarray}
\begin{array}{ll}
z = 4D\tau,  &\\ [0.2cm]
\lambda = {\Delta\over (8\pi D\tau)^{d/2}}, &
w = \left({2\tau\over \tau+t_1} \right)^{d\over 2}\,
       \exp\left[ -{({\bf x}+{\bf r}-\mbox{\boldmath $\nu$})^2
                    \over 4D(t_1+\tau)}\right],\\ [0.2cm]  
\lambda_1 = \left({\tau\over t_1}\right)^{d\over 2}, &
y = \left({2\tau\over t_1+t_2} \right)^{d\over 2}\,
       \exp\left[ -{{\bf r}^2
                    \over 4D(t_1+t_2)}\right],  \\ [0.2cm] 
\lambda_2 = 1, &
v = \left({2\tau\over \tau+t_2} \right)^{d\over 2}\,
       \exp\left[ -{({\bf x}-\mbox{\boldmath $\nu$})^2
                    \over 4D(t_2+\tau)}\right], \\ [0.2cm] 
\lambda_3 = \left({\tau\over t_2}\right)^{d\over 2},&
{\bf a}=
({\bf x}+{\bf r}-\mbox{\boldmath $\mu$})\,
     \left({2\tau\over \tau+t_1} \right)^{(d+2)\over 2}\,
       \exp\left[ -{({\bf x}+{\bf r}-\mbox{\boldmath $\mu$})^2
                    \over 4D(t_1+\tau)}\right] ,\\ [0.2cm] 
{\bf b}=(\mbox{\boldmath $\nu$ }-\mbox{\boldmath $\mu$})\,
       \exp\left[ -{(\mbox{\boldmath $\nu$}-\mbox{\boldmath $\mu$})^2
                    \over 8D\tau}\right],\quad &
{\bf c}=
({\bf x}-\mbox{\boldmath $\mu$})\,\left({2\tau\over \tau+t_2}
\right)^{(d+2)\over 2}\,
       \exp\left[ -{({\bf x}-\mbox{\boldmath $\mu$})^2
                   \over 4D(t_2+\tau)}\right].
\end{array}
\label{ojk:a}
\end{eqnarray}
\subsection{Calculation of $A$}
The elements of the matrix $A$ are calculated by constructing the
adjoint and determinant  of the inverse, since
 $A=\hbox{Adj}(A^{-1})/\hbox{det}(A^{-1})$.
Let the elements of the adjoint be defined by:
\begin{equation}
\hbox{Adj}(A^{-1})=
\pmatrix{ 
           &         &         & \vdots      &\vdots       & \vdots  \cr
 &\theta_{ij}&  &\mbox{\boldmath $\xi$}
&\mbox{\boldmath $\eta$}&\mbox{\boldmath $\zeta$} \cr
           &         &         & \vdots      & \vdots      & \vdots  \cr
\cdots&\mbox{\boldmath $\xi$}^T& \cdots    & p  &     s       &    t    \cr
\cdots&\mbox{\boldmath $\eta$}^T& \cdots   & s &     q       &    u    \cr
\cdots&\mbox{\boldmath $\zeta$}^T& \cdots  & t &     u       &    r    \cr
}.
\label{ojk:adjoint}
\end{equation}
The elements $p,q,r,\cdots ,u$ can then be calculated directly using a
 formula for the determinant of a $(d+2)\times(d+2)$ matrix of the form
\begin{equation}
B=\pmatrix{
     1      &         &  0       & \vdots      &\vdots      \cr
           & \ddots &         &{\bf a'}&{\bf b'} \cr
      0     &         &  1       & \vdots      & \vdots       \cr
\cdots&{\bf c'}^T       & \cdots    &  p'  & q'       \cr
\cdots&{\bf d'}^T     &    \cdots   &   r' &  s'       
},
\end{equation}
which is given by
\begin{equation}
\hbox{det}B = (p' -{\bf a'}.{\bf c'} )( s' -{\bf b'}.{\bf d'} ) -
 ( q'  -{\bf b'}.{\bf c'} )( r' -{\bf a'}.{\bf d' }).
\end{equation}
We derive this result by contracting the free indices in the
equation
\begin{equation}
\hbox{det}B=\sum_{ij\ldots xyz}\epsilon_{ij\ldots xyz}
B_{1\,i}B_{2\,j}\ldots B_{d+2\,z},
\label{ojk:defdetb}
\end{equation}
where $\epsilon_{ij\ldots xyz} $ is a $(d+2)$  anisotropic
tensor, which takes the value $1$ if $(ij\ldots xyz)$ is an even 
permutation of $(1\,2\,3\ldots (d+2))$, $-1$ for an odd permutation,
 and zero otherwise.

First consider the sum over the index $i\,$:
$B_{1\,i}$ is only non-zero if $i=1,d+1$ or $d+2$, ($B_{1\,1}=1$,
$B_{1\,d+1}=a'_1$ and $B_{1\,d+2}=b'_1$), which implies that equation 
(\ref{ojk:defdetb}) may be written as 
\begin{equation}
\hbox{det}B=\sum_{jk\ldots xyz}\epsilon_{1j\ldots xyz}
B_{2\,j}\ldots B_{d+2\,z}+D_1,
\label{app1:eq2}
\end{equation}
where
\begin{equation}
D_1=\sum_{jk\ldots xyz}\left(
\epsilon_{d+1jk\ldots xyz}a'_1B_{2\,j}\ldots B_{d+2\,z}+
\epsilon_{d+2jk\ldots xyz}b'_1B_{2\,j}\ldots B_{d+2\,z}
\right).
\label{d1e}
\end{equation}
We now complete the sum over $j$ in equation (\ref{d1e}) by noting
that the first term on the RHS of this expression will only be non-zero 
if $j=2$ or $j=d+2$, whereas the second term will only make a non-zero
 contribution if
$j=2$ or $j=d+1$, giving
\begin{eqnarray}
D_1&=&\sum_{k\ldots xyz}\biggl(
\epsilon_{d+1 d+2 k\ldots xyz}a'_1b'_2 B_{3\,k}\ldots B_{d+2\,z}+
\epsilon_{d+2 d+1 k\ldots xyz}b'_1a'_2 B_{3\,k}\ldots B_{d+2\,z}
\nonumber\\
&&\qquad\quad+\epsilon_{d+1 2 k\ldots xyz}a'_1 B_{3\,k}\ldots B_{d+2\,z}+
\epsilon_{d+2 2 k\ldots xyz}b'_1 B_{3\,k}\ldots B_{d+2\,z}
\biggr).
\label{app1:d1}
\end{eqnarray}
We can now complete the sum over all the remaining indices for the
first two terms on the RHS of equation (\ref{app1:d1}), since for a
non-zero contribution we must have $k=3$, $l=4,\ldots, x=d$ and 
either $y=1$ and $z=2$, or $y=2$ and $z=1$.  After completing the
sum over $k$ on the remaining RHS terms, equation  (\ref{app1:d1}) gives
\begin{eqnarray}
D_1&=&
a'_1c'_1b'_2d'_2 + b'_1d'_1a'_2c'_2 -a'_1d'_1b'_2c'_2 -
 b'_1c'_1a'_2d'_2\nonumber \\
&&+\sum_{l\ldots xyz}\biggl(
\epsilon_{d+1 2 d+2 l\ldots xyz}a'_1b'_3 B_{4\,l}\ldots B_{d+2\,z}+
\epsilon_{d+2 3 d+1 l\ldots xyz}b'_1a'_3 B_{4\,l}\ldots B_{d+2\,z}
\nonumber\\
&&\qquad\quad+\epsilon_{d+1 2 3 l\ldots xyz}a'_1 B_{4\,l}\ldots B_{d+2\,z}+
\epsilon_{d+2 2 3 l\ldots xyz}b'_1 B_{4\,l}\ldots B_{d+2\,z}
\biggr).
\label{app1:d2}
\end{eqnarray}
We can use this method repeatedly to sum over all the free indices in
equation (\ref{app1:d2}), eventually giving

\begin{eqnarray}
D_1&=&\sum_{j=2}^d\left(
a'_1c'_1b'_jd'_j + b'_1d'_1a'_jc'_j -a'_1d'_1b'_jc'_j - b'_1c'_1a'_jd'_j
 \right)\nonumber
\\&&\quad -sa'_1c'_1+q a'_1d'_1-p b'_1d'_1+r b'_1c'_1.
\label{ojk:d1}
\end{eqnarray}
Substituting equation (\ref{ojk:d1}) back into equation
 (\ref{app1:eq2}), we get:
\begin{eqnarray}
\hbox{det}B&=&\sum_{jk\ldots xyz}\epsilon_{1j\ldots xyz}
B_{2\,j}\ldots B_{d+2\,z}-sa'_1c'_1+q a'_1d'_1-p b'_1d'_1+r b'_1c'_1
\nonumber  \\
&&\quad+\sum_{j=2}^d\left(
a'_1c'_1b'_jd'_j + b'_1d'_1a'_jc'_j -a'_1d'_1b'_jc'_j - b'_1c'_1a'_jd'_j
 \right).
\end{eqnarray}
We can now repeat this entire procedure to complete the sums over the 
indices $j\ldots x$, obtaining
\begin{eqnarray}
\hbox{det}B&=&\sum_{yz}\epsilon_{12\ldots d yz}
B_{d+1\,y} B_{d+2\,z}-s{\bf a}'.{\bf c}'+q{\bf a}'.{\bf d}'
-p{\bf b}'.{\bf d}'+r{\bf b}'.{\bf c}' \nonumber \\
&&\quad+\sum_{i=1}^d\sum_{j> i}^d\left(
a'_ic'_ib'_jd'_j + b'_id'_ia'_jc'_j
 -a'_id'_ib'_jc'_j - b'_ic'_ia'_jd'_j\right).
\label{ojk:derivb}
\end{eqnarray}
Two of these terms simplify further; we find:
\begin{eqnarray}
&&\sum_{yz}\epsilon_{12\ldots d yz}
B_{d+1\,y} B_{d+2\,z}=ps-qr, \nonumber\\
&&\sum_{i=1}^d\sum_{j > i}^d\left(
a'_ic'_ib'_jd'_j + b'_id'_ia'_jc'_j -a'_id'_ib'_jc'_j
 - b'_ic'_ia'_jd'_j\right)
=({\bf a}'.{\bf c}')({\bf b}'.{\bf d}')-
({\bf a}'.{\bf d}')({\bf b}'.{\bf c}'),\nonumber
\end{eqnarray}
and therefore equation (\ref{ojk:derivb}) finally reduces to
\begin{equation}
\hbox{det}B = 
(p' -{\bf a'}.{\bf c'} )( s' -{\bf b'}.{\bf d'} ) -
 ( q'  -{\bf b'}.{\bf c'} )( r' -{\bf a'}.{\bf d' }).
\end{equation}

Having calculated the determinant of $B$, we can 
use this result to evaluate the elements $p,q,r,\cdots ,u$ of the
adjoint matrix (equation (\ref{ojk:adjoint})). We obtain 
\begin{eqnarray}
&&\,p=\left({\lambda\over z}\right)^{d+2}\,\left[
   (z\lambda_2 -{\bf b}^2 )(z\lambda_3 -{\bf c}^2 )
 -( zv - {\bf b}.{\bf c} )^2 \right] ,                \label{ojk:p}  \\  
&&\,q=\left({\lambda\over z}\right)^{d+2}\,\left[  
  (z\lambda_1 -{\bf a}^2 )(z\lambda_3 -{\bf c}^2 )-
  (zy-{\bf a}.{\bf c})^2  \right] ,                    \label{ojk:q}  \\   
&&\,r=\left({\lambda\over z}\right)^{d+2}\,\left[
  (z\lambda_1 -{\bf a}^2 )(z\lambda_2 -{\bf b}^2 )-
  (zw-{\bf a}.{\bf b})^2  \right] ,                     \label{ojk:r} \\  
&&\,s=\left({\lambda\over z}\right)^{d+2}\,\left[
  ( zv - {\bf b}.{\bf c} )(zy-{\bf a}.{\bf c})-
  (zw-{\bf a}.{\bf b})(z\lambda_3 -{\bf c}^2 ) \right],   \label{ojk:s}  \\  
&&\,t=\left({\lambda\over z}\right)^{d+2}\,\left[
  (zw-{\bf a}.{\bf b})( zv - {\bf b}.{\bf c} )-
  (z\lambda_2 -{\bf b}^2 )(zy-{\bf a}.{\bf c}) \right] ,  \label{ojk:t} \\  
&&\,u=\left({\lambda\over z}\right)^{d+2}\,\left[
  (zw-{\bf a}.{\bf b})(zy-{\bf a}.{\bf c})-
  (z\lambda_1 -{\bf a}^2 ) ( zv - {\bf b}.{\bf c}) \right]. \label{ojk:u}
\end{eqnarray}

The remaining elements of the adjoint are evaluated directly by a
component-wise expansion of the
equation $A^{-1}\,\hbox{Adj}(A^{-1})=\hbox{det}(A^{-1})\, I$,  giving
the following expressions:
\begin{eqnarray}
\mbox{\boldmath $\xi$} &=& -(p{\bf a}+s{\bf b}+t{\bf c}), \label{ojk:xi}\\ 
\mbox{\boldmath $\eta$} &=& -(s{\bf a}+q{\bf b}+u{\bf c}), \label{ojk:eta}\\ 
\mbox{\boldmath $\zeta$} &=& -(t{\bf a}+u{\bf b}+r{\bf c}), \label{ojk:zeta}\\ 
\theta_{ij} &=& {z\over\lambda}(\hbox{det}A^{-1})\delta_{ij}-
a_i\xi_j-b_i\eta_j-c_i\zeta_j. \label{ojk:theta}
\end{eqnarray}
Now that we have evaluated all the components of the adjoint, we can
evaluate the expression for the joint
probability distribution, $ P(\tilde{\bf m}) $, in terms of these variables.
Substituting equation (\ref{ojk:adjoint}) into equation (\ref{app1:pp}) and
using $A=\hbox{Adj}(A^{-1})/\hbox{det}(A^{-1})$; we find
\begin{equation}
 P(\tilde{\bf m})={1\over (2\pi)^{d+3\over 2}(\hbox{det}A^{-1})^{1\over 2}}
 \,\exp\left[-{F(\tilde{\bf m})\over 2\hbox{det}A^{-1}} \right], 
\end{equation}
where
\begin{eqnarray}
F(\tilde{\bf m})&=&\biggl(
 m'_k\theta_{kl}m'_l +2\mbox{\boldmath $\xi $}.{\bf m'}m(1) +
  2\mbox{\boldmath $\eta $}.{\bf m'}m(2)
 + 2\mbox{\boldmath $\zeta$}.{\bf m'}m(3) +pm(1)^2+qm(2)^2\nonumber\\
 &&\qquad+rm(3)^2+2sm(1)m(2)+2tm(1)m(3)+2um(2)m(3)
 \biggr).
\end{eqnarray}

\subsection{Identities Relating the Components of the Adjoint of
 $A^{-1}$ to the Determinant of $A^{-1}$ }

 In the calculation described in this paper,
there are several expressions which are frequently used to simplify
the algebra.
These are all derived directly from the component-wise 
expansion of the equation
$A^{-1}\,\hbox{Adj}(A^{-1})=\hbox{det}A^{-1}\, I$, and are listed here
for ease of reference.
\begin{eqnarray}
&&p(z\lambda_1 -{\bf a}^2 )+s (zw-{\bf a}.{\bf b})+t(zy-{\bf a}.{\bf c})=
{z\over\lambda}\hbox{det}A^{-1},\label{ojk:id1}\\
&&s(z\lambda_1 -{\bf a}^2 )+q(zw-{\bf a}.{\bf b})+u(zy-{\bf a}.{\bf
c})=0,
 \label{ojk:id2}\\
&&t(z\lambda_1 -{\bf a}^2 )+u(zw-{\bf a}.{\bf b})+r(zy-{\bf a}.{\bf c})=0 , \\
&&p(zw-{\bf a}.{\bf b})+s(z\lambda_2 -{\bf b}^2 )+t(zv-{\bf b}.{\bf c})=0,\\
&&s(zw-{\bf a}.{\bf b})+q(z\lambda_2 -{\bf b}^2 )+u(zv-{\bf b}.{\bf c})=
{z\over\lambda}\hbox{det}A^{-1},\label{ojk:id3}\\
&&t(zw-{\bf a}.{\bf b})+u(z\lambda_2 -{\bf b}^2 )+r(zv-{\bf b}.{\bf c})=0,\\
&&p(zy-{\bf a}.{\bf c})+s(zv-{\bf b}.{\bf c})+t(z\lambda_3 -{\bf c}^2
)=0,\label{ojk:id4} \\
&&s(zy-{\bf a}.{\bf c})+q(zv-{\bf b}.{\bf c})+u(z\lambda_3 -{\bf
c}^2)=0 ,\label{ojk:squ} \\
&&t(zy-{\bf a}.{\bf c})+u(zv-{\bf b}.{\bf c})+r(z\lambda_3 -{\bf c}^2 )=
{z\over\lambda}\hbox{det}A^{-1}.
\end{eqnarray}
In addition, by using the definitions of $p,q, \cdots ,u$ 
(from equations (\ref{ojk:p})-(\ref{ojk:u}))
and equations (\ref{ojk:id1}), (\ref{ojk:id2}), (\ref{ojk:id3}), 
(\ref{ojk:id4}) and
(\ref{ojk:squ}),
 we can see that
\begin{eqnarray}
&&(pq-s^2)=\left({\lambda\over z}\right)^{d+1}
           \hbox{det}A^{-1}(z\lambda_3 -{\bf c}^2), \label{ojk:pq-ss}\\
&&(qr-u^2) = \left({\lambda\over z}\right)^{d+1}
           \hbox{det}A^{-1}(z\lambda_1 -{\bf a}^2), \label{ojk:qr-uu} \\
&&(us-qt) = \left({\lambda\over z}\right)^{d+1}
           \hbox{det}A^{-1}(zy -{\bf a}.{\bf c }).  \label{ojk:us-qt}
\end{eqnarray}
 
%======================================================================
%++++++++++++++++++++++++++++++++++++++++++++++++++++++++++++++++++++++
%======================================================================

\section{Calculation of the Inverse and Determinant of  $\Omega$}
\label{ojk:App2}
\indent

In this appendix, the expressions for the determinant and inverse of the 
 matrix $\Omega$, which are required for the completion of the integrals in
equations (\ref{ojk:I1})-(\ref{ojk:I4}),  are calculated. The matrix 
$\Omega(\tilde{u},\tilde{v})$ is defined by
\begin{equation}
\Omega_{ij}(\tilde{u},\tilde{v})={1\over \hbox{det}A^{-1}}
\left(\theta_{ij}-{\eta_i\eta_j\over q}
+{(pq-s^2)\over q}h_ih_j(\tilde{u}^2-1)\right)+2\tilde{v}\delta_{ij},
\label{ojk:om}
\end{equation}
where ${\bf h}$, $\eta$ and $\theta$ are defined by equations (\ref{ojk:h}),
(\ref{ojk:xi}) and (\ref{ojk:theta}) respectively.
Substituting for ${\bf h}$, $\eta$ and $\theta$ into
  equation (\ref{ojk:om}),  and using equations 
(\ref{ojk:q}), (\ref{ojk:zeta}) and
(\ref{ojk:pq-ss})-(\ref{ojk:us-qt}), we find the expression for 
$\Omega(\tilde{u},\tilde{v})$ reduces to 
\begin{equation}
\Omega_{ij}(\tilde{u},\tilde{v})={z\over\lambda}\biggl[
\Lambda\,\delta_{ij}+\tilde{\mu}\,a_ia_j+
\tilde{\xi}\,(a_ic_j+c_ia_j)+\tilde{\nu}\,c_ic_j\biggr],
\end{equation}
where:
\begin{eqnarray}
\Lambda &=& 1+2\lambda\tilde{v}/ z ,\\
\tilde{\mu}&=&{\tilde{u}^2\over\tilde{q}}(z\lambda_3-{\bf c}^2),
\label{ojk:Tmu}\\
\tilde{\xi}&=&-{\tilde{u}^2\over\tilde{q}}(zy-{\bf a}.{\bf c}),
\label{ojk:Txi}\\
\tilde{\nu}&=&{\tilde{u}^2\over\tilde{q}}(z\lambda_1-{\bf a}^2)
+{1-\tilde{u}^2\over z\lambda_3-{\bf c}^2},\label{ojk:Tnu}\\
\tilde{q}&=&\left({z\over\lambda}\right)^{d+2}q
=(z\lambda_1 -{\bf a}^2 )(z\lambda_3 -{\bf c}^2 )-
  (zy-{\bf a}.{\bf c})^2.\label{ojk:Tq}
\end{eqnarray}
\subsection{Determinant of $\Omega(\tilde{u},\tilde{v})$ }

The determinant of the matrix $\Omega(\tilde{u},\tilde{v})$
 is calculated by evaluating the product of its
 $d$ eigenvalues.  Any vector orthogonal to both ${\bf a}$ and ${\bf c}$
 will have eigenvalue $z\Lambda/\lambda$, so we need only  
calculate the two remaining eigenvalues, which are associated with the
eigenvectors which lie 
 in the plane spanned by ${\bf a}$ and ${\bf c}$.   The eigenvalue
equation for these two may be written as
\begin{equation}
\Omega_{ij}\,(a_j+\gamma\,c_j)=\beta\,(a_i+\gamma\,c_i).
\end{equation}     
We can substitute for $\Omega(\tilde{u},\tilde{v})$
from equation (\ref{ojk:om}) and complete the sum over the index $j$;
 by equating the coefficients of 
$a_i$ and $c_i$ we then obtain two simultaneous equations for $\beta$ and 
$\gamma$, given by 
\begin{eqnarray}
\beta &=&{z\over\lambda}\biggl[ \Lambda+\tilde{\mu}{\bf a}^2
+\tilde{\xi}{\bf a.c}+\gamma(\tilde{\mu}{\bf a.c}+\tilde{\xi}{\bf
c^2})\biggr], \\ 
\beta\gamma&=&{z\over\lambda}\biggl[
\tilde{\xi}{\bf a^2}+\tilde{\nu}{\bf a.c}+\gamma(\Lambda
+\tilde{\xi}{\bf a.c}+\tilde{\nu}{\bf c^2})\biggr].
\end{eqnarray}
On eliminating $\gamma$ between these two equations, we have a
quadratic equation
in $\beta$, from which we extract the product of the two roots of the
quadratic, $\beta_\pm$.  This product is given by
\begin{equation}
\beta_+\beta_-=\left({z\over\lambda}\right)^2\Delta(\tilde{u},\tilde{v}),
\end{equation}
where
\begin{equation}
\Delta(\tilde{u},\tilde{v})=
(\Lambda+\tilde{\xi}{\bf a.c}+\tilde{\nu}{\bf c^2} )
( \Lambda+\tilde{\mu}{\bf a}^2+\tilde{\xi}{\bf a.c} )
-
(\tilde{\xi}{\bf a^2}+\tilde{\nu}{\bf a.c} )
(\tilde{\mu}{\bf a.c}+\tilde{\xi}{\bf c^2} ).
\label{ojk:product}
\end{equation}
To simplify this expression we expand the brackets on the RHS and
use equations (\ref{ojk:Tmu})-(\ref{ojk:Tq}); after some algebra, 
equation (\ref{ojk:product}) reduces to
\begin{eqnarray}
\Delta(\tilde{u},\tilde{v})&=&
\Lambda\left({\Lambda z\lambda_3-(\Lambda -1){\bf c}^2\over
z\lambda_3-{\bf c}^2}\right)-{\Lambda z\lambda_3\tilde{u}^2\over
(z\lambda_3-{\bf c}^2)}+{\Lambda z^2\tilde{u}^2\over\tilde{q}}
(\lambda_1\lambda_3-y^2)\nonumber \\
&&\quad-{(\Lambda -1) \tilde{u}^2\over\tilde{q}}
({\bf a}^2{\bf c}^2-{\bf a}.{\bf c}^2).
\label{ok:d}
\end{eqnarray}
We recall that the remaining eigenvectors all have the
eigenvalue $z\Lambda/\lambda$, and therefore the product of all the 
eigenvalues, which is equal to the  determinant, is given by 
\begin{equation}
\hbox{det}\Omega =\Lambda^{d-2}\left({z\over\lambda}\right)^d\,\Delta(\tilde{u},\tilde{v}).
\end{equation} 

\subsection{Inverse of $\Omega(\tilde{u},\tilde{v})$ }

We begin to find the inverse of  $\Omega(\tilde{u},\tilde{v})$  by
defining the variables ${\cal A}$, ${\cal B}$, ${\cal C}$ and ${\cal D}$
by the equation
\begin{equation}
\Omega^{-1}_{ij}={\lambda\over z}
({\cal A}\,\delta_{ij}+{\cal B}\,a_ia_j+{\cal C}\,(a_ic_j+c_ia_j)+
{\cal D}\,c_ic_j).
\label{oj:om}
\end{equation}
This must satisfy the identity $\Omega_{ij}\,\Omega^{-1}_{jk}=\delta_{ik}$. 
Hence, by expanding this equation and equating coefficients, we
obtain the set of simultaneous equations given below: 
\begin{eqnarray}
&O(1)\qquad &{\cal A}\Lambda = 1 ,\label{oj:0} \\
&O(a_ia_k)\qquad &{\cal A}\tilde{\mu}+
{\cal B}(\Lambda+\tilde{\mu}{\bf a}^2+
\tilde{\xi}{\bf a.c})+{\cal C}(\tilde{\mu}{\bf a.c}+
\tilde{\xi}{\bf c}^2)=0,\label{oj:1} \\
&O(a_ic_k)\qquad &{\cal A}\tilde{\xi}+
{\cal C}(\Lambda+\tilde{\mu}{\bf a}^2+\tilde{\xi}{\bf a.c})+
{\cal D}(\tilde{\mu}{\bf a.c}+\tilde{\xi}{\bf c}^2 )=0,\label{oj:2}\\
&O(c_ia_k)\qquad &{\cal A}\tilde{\xi}+
{\cal B}(\tilde{\xi}{\bf a}^2+\tilde{\nu}{\bf a.c} )+
{\cal C}(\Lambda+\tilde{\xi}{\bf a.c}+\tilde{\nu}{\bf c}^2)
=0,\label{oj:3} \\
&O(c_ic_k)\qquad &{\cal A}\tilde{\nu}+
{\cal C}(\tilde{\xi}{\bf a}^2+\tilde{\nu}{\bf a.c} )+
{\cal D}(\Lambda+\tilde{\xi}{\bf a.c}+\tilde{\nu}{\bf c}^2)
=0.\label{oj:4}
\end{eqnarray}
Since we have one more equation than we require to determine the
solutions for ${\cal A}$, ${\cal B}$, ${\cal C}$ and ${\cal D}$, we 
discard one equation and check for consistency later.
Solving equations (\ref{oj:0}), (\ref{oj:1}), (\ref{oj:2}) and
(\ref{oj:4}) simultaneously, and 
using equations (\ref{ojk:Tq}) and (\ref{ok:d}) to simplify the
results, we find, after some algebra, that:
\begin{eqnarray}
{\cal A}&=&\Lambda^{-1}, \\
{\cal B}&=-&{\tilde{u}^2\over \Lambda\tilde{q}\Delta}
                (\Lambda z\lambda_3-(\Lambda -1){\bf c}^2), \\ 
{\cal C}&=&{\tilde{u}^2\over \Lambda\tilde{q}\Delta}
                (\Lambda zy-(\Lambda -1){\bf a}.{\bf c}), \\
{\cal D}&=-&{\tilde{u}^2\over \Lambda\tilde{q}\Delta}
        (\Lambda z\lambda_1-(\Lambda -1){\bf a}^2)+
{(\tilde{u}^2-1)\over\Delta ( z\lambda_3-{\bf c}^2)}.
\end{eqnarray}
Substituting these results back into equation (\ref{oj:om}), we see that
the inverse of $\Omega(\tilde{u},\tilde{v})$ is given by
\begin{eqnarray}
\Omega^{-1}_{ij}(\tilde{u},\tilde{v})&=&{\lambda\over \Lambda z}\Biggl[
\delta_{ij}+{\Lambda(\tilde{u}^2-1)\over\Delta(z\lambda_3-{\bf c}^2)}
+{\tilde{u}^2\over\tilde{q}\Delta}
(\Lambda zy-(\Lambda -1){\bf a}.{\bf c})
(a_ic_j+c_ia_j) \nonumber \\
&&\!\!\!-{\tilde{u}^2\over\tilde{q}\Delta}\left(
(\Lambda z\lambda_3-(\Lambda -1){\bf c}^2)a_ia_j
+ (\Lambda z\lambda_1-(\Lambda -1){\bf a}^2)c_ic_j
\right)
\Biggr].
\label{ojk:fulinv}
\end{eqnarray}

This expression  may be
simplified considerably by the introduction of a new variable. If we
define ${\bf k}$ by the equation
\begin{equation}
k_i=(\Lambda z\lambda_3-(\Lambda -1){\bf c}^2)\,a_i-
(\Lambda zy-(\Lambda -1){\bf a}.{\bf c})\,c_j,
\end{equation}
then equation (\ref{ojk:fulinv}) may be rewritten as
\begin{equation}
\Omega^{-1}_{ij}={\lambda\over \Lambda z}\left(
\delta_{ij}-{c_ic_j\over (\Lambda z\lambda_3-(\Lambda -1){\bf c}^2)}
-{\tilde{u}^2\,k_ik_j\over \tilde{q}\Delta 
(\Lambda z\lambda_3-(\Lambda -1){\bf c}^2)}
\right).
\end{equation}

%======================================================================
%++++++++++++++++++++++++++++++++++++++++++++++++++++++++++++++++++++++
%======================================================================

\section{Contraction over the $i$ and $j$ Indices} 
\label{ojk:App3}
\indent
In this appendix we calculate a non-indexed expression for
\begin{equation}
\left.{\partial^2\,I_{ij}\over \partial\nu_i
 \partial\nu_j}\right|_{\mbox{\boldmath $\mu$}}=
\sum_{n=1}^4\,\left.{\partial^2\,I^{ij}_n\over \partial\nu_i
 \partial\nu_j}\right|_{\mbox{\boldmath $\mu$}}
\label{ojk:sumc}
\end{equation}
where:
\begin{eqnarray}
\left.{\partial^2\,I_1^{ij}\over \partial\nu_i \partial\nu_j}
\right|_{ \mbox{\boldmath $\mu$}
}&=&{\delta_{ij}\over d\pi 
(z\lambda_3-{\bf c}^2)[\tilde{q}\Delta(1,0)]^{1/2}} 
\left.{\partial^2\,\tilde{s}\over \partial\nu_i \partial\nu_j}
\right|_{\mbox{\boldmath $\mu$}
} \label{ojk:c2},\\
\left.{\partial^2\,I_2^{ij}\over \partial\nu_i \partial\nu_j}
\right|_{\mbox{\boldmath $\mu$}
}&=&
-{1\over \pi (z\lambda_3-{\bf c}^2)} 
\left.{\partial^2\,\tilde{s}\over \partial\nu_i \partial\nu_j}
\right|_{\mbox{\boldmath $\mu$}
}
\int_0^{\infty}d\tilde{v}\,{\Omega^{-1}_{ij}(1,\tilde{v})\over
\Lambda^{(d-2)/2}[\tilde{q}\Delta(1,\tilde{v})]^{1/2}}, \\
\left.{\partial^2\,I_3^{ij}\over \partial\nu_i \partial\nu_j}
\right|_{\mbox{\boldmath $\mu$}
}
&=& {{\bf k}(0).{\bf c}\,\delta_{ij}\over d \pi \lambda_3
 (z\lambda_3-{\bf c}^2)
[\tilde{q}\Delta(1,0)]^{1/2}}
\left.{\partial^2\,v\over \partial\nu_i \partial\nu_j}
\right|_{\mbox{\boldmath $\mu$}
},
\\
\left.{\partial^2\,I_4^{ij}\over \partial\nu_i \partial\nu_j}
\right|_{\mbox{\boldmath $\mu$}}
&=&-{z\over\pi (z\lambda_3-{\bf c}^2)}
\int_0^{\infty}{d\tilde{v}\over\Lambda^{d/2}
 [\tilde{q}\Delta(1,\tilde{v})]^{1/2}}
\, \left.{\partial^2\,v\over \partial\nu_i \partial\nu_j}
\right|_{\mbox{\boldmath $\mu$}}\nonumber\\ 
&&\times\left(
{{\bf k}(\tilde{v}).{\bf c}\,\Omega^{-1}_{ij}(1,\tilde{v})\over
 (\Lambda z\lambda_3-(\Lambda -1){\bf c}^2) }+
{\lambda (z\lambda_3 - {\bf c}^2)(k_i(\tilde{v})c_j+c_ik_j(\tilde{v}))\over
z(\Lambda z\lambda_3-(\Lambda -1){\bf c}^2)^2 }
\right),\label{ojk:c5}
\end{eqnarray}
and $\tilde{s}=(z/\lambda)^{d+2}s$, $s$ being defined by equation
(\ref{ojk:s}).

To obtain such an expression, we first evaluate the sum of the four
expressions given by equations (\ref{ojk:c2})-(\ref{ojk:c5}), and then
contract this expression over the free indices $i$ and $j$.

 To simplify these expressions we need to calculate the second 
derivative of $\tilde{s}$.  Using equations (\ref{ojk:s}) and 
(\ref{ojk:db}), we find that 
\begin{equation}
\left.{\partial^2\tilde{s}\over \partial\nu_i \partial\nu_j}
\right|_{\mbox{\boldmath $\mu$}}
=z\left((zy-{\bf a.c}) 
\left.{\partial^2v\over \partial\nu_i \partial\nu_j}
\right|_{\mbox{\boldmath$\mu$}} 
-(z\lambda_3-{\bf c}^2) 
\left.{\partial w\over \partial\nu_i \partial\nu_j}
\right|_{\mbox{\boldmath$\mu$}} 
\right).
\label{ojk:ds}
\end{equation}

After substituting for ${\bf k}$ and the second derivative
 of $\tilde{s}$ (from equations (\ref{ojk:k}) and (\ref{ojk:ds})
 respectively), into equations (\ref{ojk:c2})-(\ref{ojk:c5}), we may
combine the resulting expressions to obtain:
\begin{eqnarray}
\left.{\partial^2\,(I_1^{ij}+I_3^{ij})\over \partial\nu_i \partial\nu_j}
\right|_{\mbox{\boldmath $\mu$}} &=&
{z\delta_{ij}\over \pi d\lambda_3(\tilde{q}\Delta(1,0))^{1/2}}
\left(
y\left.{\partial^2v\over \partial\nu_i \partial\nu_j}
\right|_{\mbox{\boldmath $\mu$}}-
\lambda_3\left.{\partial^2w\over \partial\nu_i \partial\nu_j}
\right|_{\mbox{\boldmath $\mu$}}
\right)\label{ojk:I13}, \\
\left.{\partial^2\,(I_2^{ij}+I_4^{ij})\over \partial\nu_i \partial\nu_j}
\right|_{\mbox{\boldmath $\mu$}}&=&
-\int_0^{\infty}{d\tilde{v}\over
\pi \Lambda^{d/2}(\tilde{q}\Delta(1,\tilde{v}))^{1/2}}
\left[
{\lambda\,(k(\tilde{v})_ic_j+c_ik(\tilde{v})_j)
\over (\Lambda z\lambda_3-(\Lambda -1){\bf c}^2)^2}
\left.{\partial^2\,v\over \partial\nu_i \partial\nu_j}
\right|_{\mbox{\boldmath $\mu$}}
\right.\nonumber \\
&&\qquad+z\Lambda\Omega^{-1}_{ij}(1,\tilde{v})\left.
\left(
{yR\over\lambda_3}
\left.{\partial^2v\over \partial\nu_i \partial\nu_j}
\right|_{\mbox{\boldmath$\mu$}}  -
\left.{\partial^2w\over \partial\nu_i \partial\nu_j}
\right|_{\mbox{\boldmath$\mu$}}
\right)\right],
\label{ojk:I24}
\end{eqnarray}
where 
\begin{equation}
R={\lambda_3\over y}\left({\Lambda zy-(\Lambda -1){\bf a.c}\over
\Lambda z\lambda_3-(\Lambda -1){\bf c}^2}
\right)=1+{(\Lambda-1)(y{\bf c}^2-\lambda_3{\bf a}.{\bf c})\over
y(\Lambda z\lambda_3-(\Lambda -1){\bf c}^2)}.
\label{ojk:defR}
\end{equation}

Before attempting to calculate the 
sum in equation (\ref{ojk:c1}), we manipulate equation
(\ref{ojk:I24}) so that part of the $\tilde{v}$ integral can be
completed exactly.  This will simplify the algebra greatly, since  the
exactly integrable term in equation (\ref{ojk:I24}) will cancel the
contribution to the sum from equation (\ref{ojk:I13}).

We define a new variable $\Psi_{ij}$, by the equation
\begin{equation}
\Omega^{-1}_{ij}(1,\tilde{v})={\lambda\over \Lambda z}\left( 
\delta_{ij}-{\Psi_{ij}\over\tilde{q}\Delta(1,\tilde{v})}
\right);
\label{subb}
\end{equation}
on comparing this with equation (\ref{ojk:fulinv}), we find:
\begin{eqnarray}
\Psi_{ij}&=& (\Lambda z\lambda_3-(\Lambda -1){\bf c}^2)\,a_ia_j
-(\Lambda zy-(\Lambda -1){\bf a.c})\,(a_ic_j+a_jc_i)\nonumber\\
&&\qquad+(\Lambda z\lambda_1-(\Lambda -1){\bf a}^2)\,c_ic_j.
\label{ojk:psi}
\end{eqnarray}
Substituting for $R$ and $\Omega^{-1}_{ij}$ from equations (\ref{ojk:defR})
and (\ref{subb}), equation (\ref{ojk:I24}) can be written as:

\begin{eqnarray}
&&\!\!\!\!\!\!\!\!\!\!\!\!\!\!
\left.{\partial^2\,(I_2^{ij}+I_4^{ij})\over \partial\nu_i \partial\nu_j}
\right|_{\mbox{\boldmath $\mu$}}=S
-{\lambda\over\pi}\int_0^{\infty} {d\tilde{v}\over
 \Lambda^{d/2}(\tilde{q}\Delta(1,\tilde{v}))^{1/2}}\nonumber 
\\&&\times
\Biggl[
{(k(\tilde{v})_ic_j+c_ik(\tilde{v})_j)
\over (\Lambda z\lambda_3-(\Lambda -1){\bf c}^2)^2  }
\left.{\partial^2v\over \partial\nu_i \partial\nu_j}
\right|_{\mbox{\boldmath$\mu$}}
+{(\Lambda-1)(y{\bf c}^2-\lambda_3{\bf a}.{\bf c})\delta_{ij}\over
\lambda_3(\Lambda z\lambda_3-(\Lambda -1){\bf c}^2)}
\left.{\partial^2v\over \partial\nu_i \partial\nu_j}
\right|_{\mbox{\boldmath$\mu$}}\nonumber 
\\&&\quad
-{\Psi_{ij}\over \tilde{q}\Delta(1,\tilde{v})}
\left({yR\over\lambda_3}
\left.{\partial^2v\over \partial\nu_i \partial\nu_j}
\right|_{\mbox{\boldmath$\mu$}}  -
\left.{\partial^2w\over \partial\nu_i \partial\nu_j}
\right|_{\mbox{\boldmath$\mu$}}   \right)
\Biggr] ,
\label{ojk:new1}
\end{eqnarray}
where
\begin{equation}
S=-{\lambda\delta_{ij}\over\pi }
\left({y\over\lambda_3}
\left.{\partial^2v\over \partial\nu_i \partial\nu_j}
\right|_{\mbox{\boldmath$\mu$}}  -
\left.{\partial^2w\over \partial\nu_i \partial\nu_j}
\right|_{\mbox{\boldmath$\mu$}}   \right)
\int_0^{\infty} {d\tilde{v}\over
 \Lambda^{d/2}(\tilde{q}\Delta(1,\tilde{v}))^{1/2}}.
\label{ojk:ss}
\end{equation}

We now define a second new variable, $Q$, by the equation
\begin{equation}
\Delta(1,\tilde{v})=\Lambda^2\Delta(1,0)(1-Q(\tilde{v})),
\label{ok:defQ}
\end{equation}
and, using equation (\ref{ojk:delta1}), we find that
\begin{equation}
Q={\Lambda(\Lambda -1)z
(\lambda_3{\bf a^2}-2y{\bf a.c}+\lambda_1{\bf c}^2)
-(\Lambda -1)^2({\bf a}^2{\bf c}^2-{\bf a}.{\bf c}^2)\over
\Lambda^2z^2(\lambda_1\lambda_3-y^2)}\label{ojk:defQ}.
\end{equation}
 
On substituting for $\Delta(1,\tilde{v})$ from equation (\ref{ok:defQ})
into equation (\ref{ojk:ss}), and expanding the
expression $(1-Q)^{-1/2}$ using the binomial theorem, we find that
\begin{equation}
S={-\lambda\delta_{ij}\over\pi (\tilde{q}\Delta(1,0))^{1/2} }
\left({y\over\lambda_3}
\left.{\partial^2v\over \partial\nu_i \partial\nu_j}
\right|_{\mbox{\boldmath$\mu$}}  -
\left.{\partial^2w\over \partial\nu_i \partial\nu_j}
\right|_{\mbox{\boldmath$\mu$}}   \right)
\int_0^{\infty} {d\tilde{v}\over
 \Lambda^{(d+2)/2}}(1+\sum_{m=1}^{\infty}
\left( \! \begin{tabular}{c}${1\over 2}$ \\ $m$ \end{tabular} \!\!\right)
Q^m).\qquad
\label{ojk:ss1}
\end{equation}
Hence, using $\int_0^{\infty}\Lambda^{-(d+2)/2}=z/(\lambda d)$,
and inserting equation (\ref{ojk:I13}) in the expression for $S$, we
find that equation (\ref{ojk:ss1}) reduces to: 
\begin{eqnarray}
S&=&-{\lambda\delta_{ij}\over\pi (\tilde{q}\Delta(1,0))^{1/2} }
\left({y\over\lambda_3}
\left.{\partial^2v\over \partial\nu_i \partial\nu_j}
\right|_{\mbox{\boldmath$\mu$}}  -
\left.{\partial^2w\over \partial\nu_i \partial\nu_j}
\right|_{\mbox{\boldmath$\mu$}}   \right)
\int_0^{\infty} {d\tilde{v}\over
 \Lambda^{(d+2)/2}}\sum_{m=1}^{\infty}
\left( \! \begin{tabular}{c}${1\over 2}$ \\ $m$ \end{tabular} \!\!\right)
Q^m
\nonumber\\&&
\qquad
-\left.{\partial^2\,(I_1^{ij}+I_3^{ij})\over \partial\nu_i \partial\nu_j}
\right|_{\mbox{\boldmath $\mu$}}.
\label{ojk:ss2}
\end{eqnarray}

Next, substituting equation (\ref{ojk:ss2}) into equation
(\ref{ojk:new1}) and using equation (\ref{ojk:sumc}), we find that
\begin{eqnarray}
\left.{\partial^2\,I_{ij}\over \partial\nu_i
 \partial\nu_j}\right|_{\mbox{\boldmath $\mu$}}&=&
-{\lambda\delta_{ij}\over\pi (\tilde{q}\Delta(1,0))^{1/2}}
\int_0^{\infty} {d\tilde{v}\over
 \Lambda^{(d+2)/2}}\sum_{m=1}^{\infty}
\left( \! \begin{tabular}{c}${1\over 2}$ \\ $m$ \end{tabular} \!\!\right)
Q^m\nonumber \\
&& -{\lambda\over\pi (\tilde{q}\Delta(1,0))^{1/2}}
\int_0^{\infty} {d\tilde{v}\over \Lambda^{(d+2)/2}(1-Q)^{1/2}}
\nonumber \\
&&\!\!\!\!\times
\Biggl[
{(\Lambda-1)(y{\bf c}^2-\lambda_3{\bf a}.{\bf c})\delta_{ij}\over
\lambda_3(\Lambda z\lambda_3-(\Lambda -1){\bf c}^2)}
\left.{\partial^2v\over \partial\nu_i \partial\nu_j}
\right|_{\mbox{\boldmath$\mu$}} 
+{(k(\tilde{v})_ic_j+c_ik(\tilde{v})_j)
\over (\Lambda z\lambda_3-(\Lambda -1){\bf c}^2)^2  }
\left.{\partial^2v\over \partial\nu_i \partial\nu_j}
\right|_{\mbox{\boldmath$\mu$}}
\nonumber \\
&&\qquad
-{\Psi_{ij}\over \Lambda^2\tilde{q}\Delta(1,0)(1-Q)}
\left({yR\over\lambda_3}
\left.{\partial^2v\over \partial\nu_i \partial\nu_j}
\right|_{\mbox{\boldmath$\mu$}}  -
\left.{\partial^2w\over \partial\nu_i \partial\nu_j}
\right|_{\mbox{\boldmath$\mu$}}   \right)
\Biggr].
\label{csta}
\end{eqnarray}

Before we start the contraction of equation (\ref{csta}) over the indices
$i$ and $j$ it is convenient to derive some useful results.

First, we calculate the derivatives of $v$ and $w$ (from equations
(\ref{ojk:a})), these are given by: 
\begin{eqnarray}
\left. {\partial^2v\over \partial\nu_i \partial\nu_j}
\right|_{\mbox{\boldmath $\mu $}} & =&
\left[{c_ic_j\over z^2 v}-{v\delta_{ij}\over 2D(t_2+\tau)} 
\right]_{\mbox{\boldmath $\mu $}}, \\
\left. {\partial^2w\over \partial\nu_i \partial\nu_j}
\right|_{\mbox{\boldmath $\mu $}} & =&
\left[{a_ia_j\over z^2 w}-{w\delta_{ij}\over 2D(t_1+\tau)}
\right]_{\mbox{\boldmath $\mu $}}.
\label{ojk:duv}
\end{eqnarray}
(From this point on, we will use the notation;
 $v_{\mu}$ and  $w_{\mu}$ to represent $v$ and $w$
evaluated $\mbox{\boldmath $\nu $}=\mbox{\boldmath $\mu
$}$).

Using these equations together with equations (\ref{ojk:k}) and
 (\ref{ojk:psi}), we  evaluate the following expressions:
\begin{eqnarray}
&&\delta_{ij}\left.{\partial^2\,v\over \partial\nu_i \partial\nu_j}
\right|_{ \mbox{\boldmath $\mu$}}=
\left({{\bf c}^2\over z^2 v_\mu}-{v_\mu d\over 2D(t_2+\tau)}\right)
,  \label{ojk:du} \\
&&\delta_{ij}\left.{\partial^2\,w\over \partial\nu_i \partial\nu_j}
\right|_{ \mbox{\boldmath $\mu$}}=
\left({{\bf a}^2\over z^2 w_\mu}-{w_\mu d\over 2D(t_1+\tau)}\right)
,  \label{ojk:dv} \\
&&(k(\tilde{v})_ic_j+c_ik(\tilde{v})_j)
\left.{\partial^2\,v\over \partial\nu_i \partial\nu_j}
\right|_{ \mbox{\boldmath $\mu$}}=
2z\Lambda(\lambda_3{\bf a}.{\bf c}-y{\bf c}^2)
%\nonumber\\&&\qquad\qquad\qquad\qquad\qquad\qquad\times
\left(
{{\bf c}^2\over z^2v_{\mu}}-{v_{\mu}\over 2D(t_2+\tau)}
\right), \qquad\qquad \label{ojk:sim3} \\
&&\Psi_{ij}(\tilde{v})\,\delta_{ij}
=\Lambda z (\lambda_3{\bf a}^2-2y{\bf a.c}+\lambda_1{\bf c}^2)
-2(\Lambda -1) ({\bf a}^2{\bf c}^2-{\bf a.c}^2),
\label{ojk:psid} \\
&&\Psi_{ij} \left.{\partial^2\,v\over \partial\nu_i \partial\nu_j}
\right|_{ \mbox{\boldmath $\mu$}}=
\left({{\bf c}^2\over z^2v_{\mu}}-{v_{\mu}\over 2D(t_2+\tau)}\right)
\Psi_{ij}\,\delta_{ij}
\nonumber \\ && \qquad\qquad\qquad\qquad\qquad\qquad
-{(\Lambda z\lambda_3-(\Lambda -1){\bf c}^2)\over z^2v_{\mu}}
({\bf a}^2{\bf c}^2-{\bf a.c}^2) ,\label{ojk:sim1} 
\\
&&\Psi_{ij}\left.{\partial^2\,w\over \partial\nu_i \partial\nu_j}
\right|_{ \mbox{\boldmath $\mu$}}= 
 \left({{\bf a}^2\over z^2w_{\mu}}-{w_{\mu}\over 2D(t_1+\tau)}\right) 
\Psi_{ij}\,\delta_{ij}\nonumber \\ && \qquad\qquad\qquad\qquad\qquad\qquad
-{(\Lambda z\lambda_1-(\Lambda -1){\bf a}^2)\over z^2w_{\mu}}
({\bf a}^2{\bf c}^2-{\bf a.c}^2). 
\label{ojk:sim2}
\end{eqnarray}
Finally, we can complete the contraction over $i$ and
$j$ by substituting equations (\ref{ojk:du})-(\ref{ojk:sim2}),
 together with  (\ref{ojk:defR}) and (\ref{ojk:defQ})) into equation
 (\ref{csta}) to give:
\begin{equation}
\left.{\partial^2\,I_{ij}\over \partial\nu_i
 \partial\nu_j}\right|_{\mbox{\boldmath $\mu$}}=-{\lambda\over z\pi}
\int_0^{\infty}d\tilde{v}\,\sum_{n=1}^6T_n,
\end{equation}
where:
\begin{eqnarray}
T_1&=&
{\sum_{m=1}^{\infty}
\left( \! \begin{tabular}{c}${1\over 2}$ \\ $m$ \end{tabular} \!\!\right)
 Q^m
\over (\lambda_1\lambda_3-y^2)^{1/2}\Lambda^{(d+2)/2}}
\nonumber \\ &&\times
\left({y\over\lambda_3}
\left({{\bf c}^2\over z^2v_{\mu}}-{v_{\mu}d\over 2D(t_2+\tau)}\right)
- 
\left({{\bf a}^2\over z^2w_{\mu}}-{w_{\mu}d\over 2D(t_1+\tau)}\right)
\right),
\label{ojk:t1} \\
T_2&=&{(\Lambda-1)( y{\bf c}^2- \lambda_3{\bf a.c})
\over \Lambda^{(d+2)/2}\lambda_3
(\lambda_1\lambda_3-y^2)^{1/2}(1-Q)^{1/2}
(\Lambda z\lambda_3-(\Lambda -1){\bf c}^2)}
\nonumber \\ &&\times
\left( {{\bf c}^2\over z^2v_{\mu}}-{v_{\mu}d\over 2D(t_2+\tau)}\right),
\label{ojk:t2} \\
T_3&=&{2z(\lambda_3{\bf a.c}-y{\bf c}^2)\over
 \Lambda^{d/2}(\lambda_1\lambda_3-y^2)^{1/2}(1-Q)^{1/2}
(\Lambda z\lambda_3-(\Lambda -1){\bf c}^2)^2}
\nonumber \\&& \times
\left({{\bf c}^2\over z^2v_{\mu}}-{v_{\mu}\over 2D(t_2+\tau)}  \right),
\label{ojk:t3} \\
T_4&=&{z^{-4}({\bf a}^2{\bf c}^2-{\bf a.c}^2)\over
\Lambda^{(d+6)/2}(\lambda_1\lambda_3-y^2)^{3/2}(1-Q)^{3/2}}
\nonumber \\&& \times
\left({(\Lambda zy-(\Lambda -1){\bf a.c})\over  v_{\mu} }
-{(\Lambda z\lambda_1-(\Lambda -1){\bf a}^2)\over  w_{\mu} }
\right),
\label{ojk:t4} \\
T_5&=&
{-(\lambda_3{\bf a}^2-2y{\bf a}.{\bf c}+\lambda_1{\bf c}^2)\over
\Lambda^{(d+4)/2}z(\lambda_1\lambda_3-y^2)^{3/2}(1-Q)^{3/2}}
\nonumber \\&& \times
\Biggl({yR\over\lambda_3}
\left({{\bf c}^2\over z^2v_{\mu}}-{v_{\mu}\over 2D(t_2+\tau)}\right)
- \left({{\bf a}^2\over z^2w_{\mu}}-{w_{\mu}\over 2D(t_1+\tau)}\right)
\Biggr),
\label{ojk:t5} \\
T_6&=&
{2(\Lambda-1)({\bf a}^2{\bf c}^2-{\bf a.c}^2)\over
\Lambda^{(d+6)/2}z^2(\lambda_1\lambda_3-y^2)^{3/2}(1-Q)^{3/2}}
\nonumber \\&& \times
\Biggl({yR\over\lambda_3}
\left({{\bf c}^2\over z^2v_{\mu}}-{v_{\mu}\over 2D(t_2+\tau)}\right)
- \left({{\bf a}^2\over z^2w_{\mu}}-{w_{\mu}\over 2D(t_1+\tau)}\right)
\Biggl)
\label{ojk:t6}.
\end{eqnarray}
%

%======================================================================
%++++++++++++++++++++++++++++++++++++++++++++++++++++++++++++++++++++++
%======================================================================

\section{}
\label{ojk:App4}
\subsection{The Large-$d$ Behavior of the $T_n$}

In this section we will demonstrate that the large-$d$ behavior of
the  terms, $T_i$, which are defined by equations 
(\ref{ojk:t1})-(\ref{ojk:t6}), is 
controlled, and hence that we are justified in applying the method of
steepest descents to the $\hat{\mu}$ and $\psi$ integrals 
in equation (\ref{ojk:corrsub}). 

To simplify the algebra within this section we define several new
 variables:
\begin{eqnarray}
& f_1=(1-x/d),      
& \gamma=\left({4t_1t_2\over (t_1+t_2)^2} \right)^{d/4}
\exp\left({-r^2\over 4D(t_1+t_2)}\right),     \nonumber \\
& f_2=(1-x/2d),      
& E=f_2^{-d+2\over 2} \exp\left( -{x\hat{\mbox{\boldmath $\mu$}}^2\over
  8Df_2}\right),     \nonumber \\
& f_3=(1-{t_2\over t_1+t_2}x/d),\quad     
& E_s=f_3^{-d+2\over 2}
\exp\left({{\bf r}^2\over 4D(t_1+t_2)}
-{({\bf r}-(xt_2)^{1\over 2}\hat{\mbox{\boldmath $\mu$}})^2
 \over 4D(t_1+t_2)f_3 }
\right). \nonumber    
\end{eqnarray}
Therefore, the set of variables given by equations (\ref{ojk:a}), 
with which the terms $T_i$ are defined, may be rewritten as:
\begin{eqnarray}
\begin{array}{ll}
z = 4Dt_2f_1, &
w_{\mu} = \gamma f_3 f_1^{d/2}
\left({t_2\over t_1} \right)^{d/4}\, E_s,      \\ [0.2cm]
\lambda_1 = \left({t_2\over t_1}\right)^{d/2}f_1^{d/2},\qquad &
y = \left({t_2\over t_1} \right)^{d/4} f_1^{d/2}\gamma,   \\ [0.2cm]
\lambda_3 = f_1^{d/2}, &
{\bf a}= {2t_2\gamma\over t_1+t_2}
f_1^{(d+2)/2}
\left({t_2\over t_1} \right)^{d/4}
\left[{\bf r}-(xt_2)^{1\over 2}\hat{\mbox{\boldmath $\mu$}}
\right]\,E_s, \\[0.2cm]
 v_{\mu} = f_2f_1^{d/2}\,E, &
{\bf c}=-(xt_2)^{1\over 2}f_1^{(d+2)/2}
 \hat{\mbox{\boldmath $\mu$}}\,E.
\end{array}
\label{ojk:newvar}
\end{eqnarray}

To investigate the large-$d$ behavior of the $T_n$, we substitute
these variables back into equations (\ref{ojk:t1})-(\ref{ojk:t6})
to obtain:
\begin{eqnarray}
T_1&=&
{\gamma \sum_{m=1}^\infty 
\left( \! \begin{tabular}{c}${1\over 2}$ \\ $m$ \end{tabular} \!\!\right)
Q^m \over 2D\Lambda^{(d+2)/2} (1-\gamma^2)^{1/2}}
\nonumber \\ && \times
\left[{E\over 2t_2}\left({x\hat{\mbox{\boldmath $\mu$}}^2\over 4Df_2 }
-d\right)
-{E_s\over t_1+t_2}\left(
{({\bf r}-(xt_2)^{1\over 2}
\hat{\mbox{\boldmath $\mu$}})^2\over 2Df_3(t_1+t_2) } -d\right)
\right], \label{ojk:tt1}\\ [0.2cm]
T_2&=&{-\gamma  (\Lambda-1)f_1^{(d+2)/2}E^2\over
4Dt_2(t_1+t_2) \Lambda^{(d+2)/2}(1-\gamma^2)^{1/2}(1-Q)^{1/2}}
\left({x\hat{\mbox{\boldmath $\mu$}}^2\over 4Df_2 }-d \right)
\nonumber\\
&&\times
{(x\hat{\mbox{\boldmath $\mu$}}^2[2t_2E_s-(t_1+t_2)E]-
2(xt_2)^{1\over 2}\hat{\mbox{\boldmath $\mu$}}.{\bf r}E_s)
\over
4D\Lambda -
(\Lambda-1)f_1^{(d+2)/2}x\hat{\mbox{\boldmath $\mu$}}^2E^2},
\label{ojk:tt2} \\[0.2cm]
T_3&=&{2\gamma f_1^{(d+2)/2}E^2\over
\Lambda^{d/2} t_2(t_1+t_2)(1-\gamma^2)^{1/2}(1-Q)^{1/2}}
\left({x\hat{\mbox{\boldmath $\mu$}}^2\over 4Df_2 }-1 \right)
\nonumber\\&&\times
{(x\hat{\mbox{\boldmath $\mu$}}^2[2t_2E_s-(t_1+t_2)E]-
2(xt_2)^{1\over 2}\hat{\mbox{\boldmath $\mu$}}.{\bf r}E_s  )
\over(4D\Lambda-
(\Lambda-1)f_1^{(d+2)/2} x\hat{\mbox{\boldmath $\mu$}}^2E^2
)^2 },\label{ojk:tt3} \\[0.2cm]
T_4&=&{xf_1^{(d+2)/2}EE_s\,
(\hat{\mbox{\boldmath $\mu$}}^2{\bf r}^2-
\hat{\mbox{\boldmath $\mu$}}.{\bf r}^2) \over
16D^4 \Lambda^{(d+6)/2} (t_1+t_2)^2(1-\gamma^2)^{3/2}(1-Q)^{3/2}}
\nonumber \\
&& \times\Biggl[{\gamma^3E_s\over f_2}\left(
D\Lambda-
{(\Lambda -1)f_1^{(d+2)/2}EE_s\over2(t_1+t_2)}
( xt_2\hat{\mbox{\boldmath $\mu$}}^2
-(xt_2)^{1\over 2}\hat{\mbox{\boldmath $\mu$}}.{\bf r} )
 \right)
\nonumber\\
&&\qquad
-{\gamma E\over f_3}\left(
D\Lambda-
{\gamma^2(\Lambda -1)t_2f_1^{(d+2)/2}E_s^2\over(t_1+t_2)^2}
({\bf r}-(xt_2)^{1\over 2}\hat{\mbox{\boldmath $\mu$}})^2
 \right)
\Biggr],\label{ojk:tt4}\\[0.2cm]
T_5&=&{-\gamma (1-\gamma^2)^{-3/2}f_1^{(d+2)/2} 
\over
8D^2\Lambda^{(d+4)/2} (1-Q)^{3/2}}
\left[{RE\over 2t_2}
\left({x\hat{\mbox{\boldmath $\mu$}}^2\over 4Df_2 }-1 \right)
-{E_s\over t_1+t_2}
\left({({\bf r}-(xt_2)^{1\over 2}\hat{\mbox{\boldmath $\mu$}})^2
\over 2D(t_1+t_2)f_3 }-1 \right)\right]\nonumber \\
&&\times\biggl[ xE^2\hat{\mbox{\boldmath $\mu$}}^2+
{4t_2\gamma^2 E_s^2\over (t_1+t_2)^2}
  ({\bf r}-(xt_2)^{1\over 2}\hat{\mbox{\boldmath $\mu$}})^2
%\nonumber \\&&
-{4\gamma^2EE_s\over (t_1+t_2)}
(xt_2\hat{\mbox{\boldmath $\mu$}}^2-
(xt_2)^{1\over 2}\hat{\mbox{\boldmath $\mu$}}.{\bf r})
\biggr],\label{ojk:tt5}\\[0.2cm]
T_6&=&{(\Lambda-1)xt_2\gamma^3f_1^{d+2}E^2E_s^2 (\hat{\mbox{\boldmath $\mu$}}^2{\bf r}^2-
\hat{\mbox{\boldmath $\mu$}}.{\bf r}^2) 
\over
4D^3(t_1+t_2)^2 \Lambda^{(d+6)/2}
 (1-\gamma^2)^{3/2}(1-Q)^{3/2}}\nonumber \\
&&\times
\left[{RE\over 2t_2}\left({x\hat{\mbox{\boldmath $\mu$}}^2\over 4Df_2 }
-1\right)-
{E_s\over (t_1+t_2)}\left(
{({\bf r}-(xt_2)^{1\over 2}\hat{\mbox{\boldmath $\mu$}})^2\over 2D(t_1+t_2)f_3 } -1\right)
\right],\label{ojk:tt6}
\end{eqnarray}
where: 
\begin{eqnarray}
Q&=&{(\Lambda -1)f_1^{(d+2)/2}\over 4\Lambda D(1-\gamma^2)}
\Biggl[
{4t_2\gamma^2 E_s^2\over (t_1+t_2)^2}
  ({\bf r}-(xt_2)^{1\over 2}\hat{\mbox{\boldmath $\mu$}})^2
-{4\gamma^2EE_s\over (t_1+t_2)}
(xt_2\hat{\mbox{\boldmath $\mu$}}^2-
(xt_2)^{1\over 2}\hat{\mbox{\boldmath $\mu$}}.{\bf r})
\nonumber\\ 
&& \qquad +xE^2\hat{\mbox{\boldmath $\mu$}}^2
-{xt_2(\Lambda -1)f_1^{(d+2)/2}\over \Lambda D}\left(
{\gamma E E_s\over t_1+t_2}\right)^2
(\mbox{\boldmath $\mu$}^2{\bf r}^2-\hat{\mbox{\boldmath $\mu$}}.{\bf
r}^2)
\Biggr],
\label{ojk:Q}\\
R&=&
1+{(\Lambda -1)Ef_1^{(d+2)/2}
(x\hat{\mbox{\boldmath $\mu$}}^2[(t_1+t_2)E-2t_2E_s]
+2E_s(xt_2)^{1\over 2}\hat{\mbox{\boldmath $\mu$}}.{\bf r})
\over(t_1+t_2)
(4D\Lambda-(\Lambda -1)f_1^{(d+2)/2}x\hat{\mbox{\boldmath $\mu$}}^2E^2)},
 \label{ojk:R}
\end{eqnarray}
 $\Lambda=1+2\tilde{w}/d$, and  
$\left( \! \begin{tabular}{c}${1\over 2}$ \\ $m$ \end{tabular}
\!\!\right)$ are binomial coefficients.

Although the above expressions are rather complicated, we are only
interested in determining whether the large-$d$ behavior in each
case is bounded. 
 We notice that all the factors 
$(t_2/t_1)^{d/4}$ cancel, and the only terms which have a
$d$-dependent exponent are
$f_1^{d/2}$, $f_2^{d/2}$, $f_3^{d/2}$ and $\Lambda^{d/2}$. 
  However, the large-$d$ limit of each of these expressions 
 is independent of $d$.  We have:
\begin{eqnarray}
\begin{array}{ll}
\lim_{d\to\infty}f_1^{d/2}=\exp(-{x\over 2}),
&\lim_{d\to\infty}f_2^{d/2}=\exp(-{x\over 4}), \cr
\lim_{d\to\infty}f_3^{d/2}=\exp(-{t_2x\over 2(t_1+t_2)}),\qquad
&\lim_{d\to\infty}\Lambda^{-d/2}=\exp(-\tilde{w}).
\end{array}
\end{eqnarray}
At large $d$ therefore, all the terms $T_n$ in equation
(\ref{ojk:corrsub}), are dominated by the exponential factor 
$\exp[-dg(x)]$,  and
hence we may complete the integral using the method of steepest descents.

\subsection{The $1/d$ Expansion}

In this section we evaluate  the leading order term in the $1/d$
expansion of each of the expressions for $T_n$ (given by equations
(\ref{ojk:tt1})-(\ref{ojk:tt6})) evaluated at $\hat{\mbox{\boldmath
$\mu$}}^2=2D$ and $\psi=\pi /2$, where
$\psi$ is the angle between ${\bf r}$ and $\hat{\mbox{\boldmath $\mu$}}$.

Since the final part of this calculation requires the integration of
these terms over the variables $x$ and $\tilde{w}$, it is important to
check that these integrals do not alter the $d$-dependence of
higher orders in the expansion; this ensures that we have calculated
the entire leading order contribution.
If we examine the expression for each $T_n$ in turn, we see that, in the
large-$d$ limit, every order in the $1/d$ expansion will have an
exponential factor with a negative $x$ and $\tilde{w}$ exponent.  The
presence of these exponential factors ensures that the $x$ and
$\tilde{w}$ integrals will not alter the $d$-dependence at any order
in the expansion, so we only need to calculate the leading order terms. 

We first evaluate the variables $E$, $E_s$, $R$ and $Q$ at the
 position of the minimum which controls the value of the integral in 
equation (\ref{ojk:corrsub}) ($\hat{\mu}=\sqrt{2D},\psi=\pi /2$); then
 expanding to leading order in $1/d$ gives $E=1$, $E_s=1$, $R=1$ and
\begin{equation}
Q={\tilde{w}\exp(-x/2)\over d(1-\gamma^2)}
\left(x\left(1-{4t_1t_2\gamma^2\over(t_1+t_2)^2}\right)
+{2t_2\gamma^2{\bf r}^2\over D(t_1+t_2)^2}
\right).
\end{equation}

Using these results we can now calculate the terms up to $O(1)$ in the
 expansion of each expression for $T_n$ (equations
 (\ref{ojk:tt1})-(\ref{ojk:tt6})); these are given by:
\begin{eqnarray}
T_1&=&{\gamma(t_2-t_1)\exp(-x/2)\,\tilde{w}\exp(-\tilde{w})\over
8Dt_2(t_1+t_2)(1-\gamma^2)^{3/2}}
\nonumber \\&&\times
\left(x\left(1-{4t_1t_2\gamma^2\over(t_1+t_2)^2}\right)
+{2t_2\gamma^2{\bf r}^2\over D(t_1+t_2)^2}
\right), \\ [0.2cm]
T_2&=&{\gamma (t_2-t_1)x\exp(-x/2)\,\tilde{w}\exp(-\tilde{w})\over
4 Dt_2(t_1+t_2)(1-\gamma^2)^{1/2}},  \\[0.2cm]
T_3&=&{\gamma (t_2-t_1)x(x-2)\exp(-x/2)\,\exp(-\tilde{w})\over 
8 Dt_2(t_1+t_2)(1-\gamma^2)^{1/2}}, \\[0.2cm]
T_4&=&{-\gamma{\bf r}^2x\exp(-x/2)\exp(-\tilde{w})\over 
8 D^2(t_1+t_2)^2(1-\gamma^2)^{1/2}},  \\[0.2cm]
T_5&=&{-\gamma\exp(-x/2)\exp(-\tilde{w})\over 8Dt_2(1-\gamma^2)^{3/2}}
\left(x\left(1-{4t_1t_2\gamma^2\over(t_1+t_2)^2}\right)
+{2t_2\gamma^2{\bf r}^2\over D(t_1+t_2)^2}
\right)\nonumber \\
&&\qquad\times \left[ {t_2-t_1\over t_1+t_2} -{t_2{\bf r}^2\over D(t_1+t_2)^2}
+{x\over 2}\left(1-{4t_2^2\over (t_1+t_2)^2}\right)\right], \\[0.2cm]
T_6&=&O(1/d). 
\end{eqnarray}

%+++++++++++++++++++++++++++++++++++++++++++++++++++++++++++
%+++++++++++++++++++++++++++++++++++++++++++++++++++++++++++

\end{document}